\documentstyle[12pt]{article}
\begin{document}
\newcommand{\la}{\langle}
\newcommand{\ra}{\rangle}
\renewcommand{\d}{\partial}
\newcommand{\beq}{\begin{eqnarray}}
\newcommand{\eeq}{\end{eqnarray}}
\newcommand{\sbeq}{\begin{subeqnarray}}
\newcommand{\seeq}{\end{subeqnarray}}
\newcommand{\eps}{\epsilon}
\newcommand{\cl}{\centerline}
\newcommand{\btem}{\bibitem}
\newcommand{\YO}{Y. Oono}
\newcommand{\NG}{N. Goldenfeld}
\newcommand{\bfg}{\mbox{{\boldmath $g$}}}
\newcommand{\bfh}{\mbox{{\boldmath $h$}}}
\newcommand{\bfs}{\mbox{{\boldmath $s$}}}
\newcommand{\bfG}{\mbox{{\boldmath $G$}}}
\newcommand{\bfu}{\mbox{{\boldmath $u$}}}
\newcommand{\bfe}{\mbox{{\boldmath $e$}}}
\newcommand{\bfU}{\mbox{{\boldmath $U$}}}
\newcommand{\bfW}{\mbox{{\boldmath $W$}}}
\newcommand{\bfX}{\mbox{{\boldmath $X$}}}
\newcommand{\bfV}{\mbox{{\boldmath $V$}}}
\newcommand{\bfF}{\mbox{{\boldmath $F$}}}
\newcommand{\bfP}{\mbox{{\boldmath $P$}}}
\newcommand{\bfC}{\mbox{{\boldmath $C$}}}
\newcommand{\bfR}{\mbox{{\boldmath $R$}}}
\newcommand{\bfrho}{\mbox{{\boldmath $\rho$}}}
\newcommand{\brho}{\bfrho}
\newcommand{\e}{{\rm e}}
\newcommand{\half}{{1\over 2}}
\newcommand{\PRL}{Phys. Rev. Lett.}
\newcommand{\bmR}{\mbox{\boldmath$R$}}
\newcommand{\bmC}{\mbox{\boldmath$C$}}
\newcommand{\bmb}{\mbox{\boldmath$b$}}
\newcommand{\bmh}{\mbox{\boldmath$h$}}
\newcommand{\bmk}{\mbox{\boldmath$k$}}
\newcommand{\bmH}{\mbox{\boldmath$H$}}
\newcommand{\bmZ}{\mbox{\boldmath$Z$}}
\newcommand{\bmN}{\mbox{\boldmath$N$}}
\newcommand{\bma}{\mbox{\boldmath$a$}}
\newcommand{\bmu}{\mbox{\boldmath$u$}}
\newcommand{\bmv}{\mbox{\boldmath$v$}}
\newcommand{\bmw}{\mbox{\boldmath$w$}}
\newcommand{\bm}[1]{\mbox{\boldmath$#1$}}
\newcommand{\bmzero}{\bm{0}}
\newcommand{\bphi}{{\bf \phi }}
\newcommand{\mod}{{\rm mod}}
\newcommand{\ep}{\epsilon}
\newcommand{\pa}{\partial}
\newcommand{\rf}[1]{(\ref{#1})}
\newcommand{\dfrac}[2]{\displaystyle \frac{#1}{#2}}
\newcommand{\dint}{\displaystyle \int}
\newcommand{\dsum}{\displaystyle \sum}
\newcommand{\dsup}{\displaystyle \sup}
\newcommand{\dinf}{\displaystyle \inf}
\newcommand{\dlim}{\displaystyle \lim}
\newcommand{\dmax}{\displaystyle \max}
\newcommand{\dmin}{\displaystyle \min}
\newcommand{\dcup}{\displaystyle \cup}
\newcommand{\spvii}{\; \; \; \; \; \; \; }
\newcommand{\lng}{\left\langle  \right.}
\newcommand{\rng}{\left.  \right\rangle}
\newcommand{\vep}{\varepsilon}
\newcommand{\ode}[2]{\displaystyle \frac{d #1}{d #2}}
\newcommand{\pde}[2]{\displaystyle \frac{\pa #1}{\pa #2}}
\newcommand{\calL}{{\cal L}}
\newcommand{\calM}{{\cal M}}
\newcommand{\ve}{\varepsilon}
\newcommand{\ovl}[1]{\overline{#1}}
\newcommand{\ovlbh}{\overline{\bmh}}
\newcommand{\ovlh}{\overline{h}}
\newcommand{\ovldel}{\overline{\delta}}
\newcommand{\ovlcM}{\widehat{\calM }}
\newcommand{\unl}[1]{\underline{#1}}
\newcommand{\unlambda}{\underline{\lambda}}
\newcommand{\unLambda}{\underline{\Lambda}}
\newcommand{\TPi}{\widetilde{\Pi}}
\newcommand{\TP}{\widetilde{P}}
\newcommand{\Tbh}{\widehat{\bmh}}
\newcommand{\Th}{\widehat{h}}
\newcommand{\TF}{\widetilde{F}}
\newcommand{\Tg}{\tilde{g}}
\newcommand{\TG}{\widetilde{G}}
\newcommand{\TU}{\widetilde{U}}
\newcommand{\Tw}{\tilde{w}}
\newcommand{\TS}{\tilde{S}}
\newcommand{\Tsgm}{\widehat{\sigma}}
\newcommand{\Tbmv}{\tilde{\bmv}}
\newcommand{\TW}{\widetilde{W}}
\newcommand{\TiH}{\widetilde{H}}
\newcommand{\TbmH}{\widetilde{\bmH}}
\newcommand{\Hv}{\hat{v}}
\newcommand{\Epe}{E ^{\perp}}
\begin{center}
{\Large  Renormalization-group Method for Reduction of Evolution Equations;\ \ 
invariant manifolds and envelopes
}\\ 
\vspace{1cm}
 SHIN-ICHIRO EI$^{1}$,  KAZUYUKI FUJII$^2$ and 
 TEIJI KUNIHIRO$^3$\\ 
 \vspace{.5cm}
$^1$Graduate School of Integrated Science,
 Yokohama City University, \\ 
 Yokohama 236-0027 Japan,\\ 
$^2$Department of Mathematical Sciences, Yokohama City University, \\ 
 Yokohama 236-0027 Japan,\\ 
$^3$ Faculty of Science and Technology, Ryukoku University,\\
Otsu, 520-2194 Japan
\end{center}
\begin{abstract}
The renormalization group (RG) method as a powerful tool
 for reduction of evolution equations is formulated 
in terms of the notion of invariant manifolds.
We start with  derivation of an exact RG equation which is analogous to
 the Wilsonian RG equations in statistical physics and quantum field theory.
It is clarified that the perturbative RG method constructs
 invariant manifolds  successively as the initial value of 
evolution equations, thereby the meaning to set $t_0=t$ is naturally 
understood where $t_0$ is the arbitrary initial time. We show that 
the integral constants in the unperturbative 
solution constitutes natural coordinates of the invariant manifold
when the linear operator $A$ in the evolution equation is semi-simple,
i.e., diagonalizable;
when $A$ is not semi-simple and has a Jordan cell,
 a slight modification is necessary because 
the dimension of the invariant manifold 
is increased  by the perturbation.
The RG equation determines the slow motion
of the would-be integral constants in the unperturbative solution
 on the invariant manifold.
 We present the mechanical procedure to construct the perturbative
 solutions hence the initial values 
with which the RG equation gives meaningful results.
The underlying structure of the reduction by the
 RG method as formulated in the present work 
turns out to completely fit to the universal one
  elucidated  by Kuramoto some years ago. 
We indicate that the reduction procedure of evolution equations 
 has a good correspondence with the renormalization procedure
 in quantum field theory; the counter part of 
the universal structure of reduction elucidated by Kuramoto
 may be the Polchinski's theorem for renormalizable field theories.
 We apply the method to interface dynamics such as 
 kink-anti-kink  and soliton-soliton interactions in the latter of which 
 a linear operator having a  Jordan-cell structure appears.
\end{abstract}
\vspace{1cm}
67 pages including the firt two pages; no figures, no tables.
\newpage 
The Proposed running head:\\ 
Renormalization-group Method 
\\ 
Correspondence should be sent to\\ 
Prof. T. Kunihiro,\\ 
 Faculty of Science and Technology, Ryukoku University,\\
Otsu, 520-2194 Japan\\ 
Tel. (+81)-77-543-7501 \\ 
Fax. (+81)-77-543-7524 \\ 
e-mail  kunihiro@rins.ryukoku.ac.jp
\newpage 
\section{Introduction}
\setcounter{equation}{0}
\renewcommand{\theequation}{\thesection.\arabic{equation}}

There is an ever growing interest in the renormalization groups (RG) 
\cite{rg,JZ,shirkov} in various fields of science and mathematical 
physics since the work of Wilson\cite{wilson,weg,pol}.
The essence of the RG in quantum field theory (QFT) 
and statistical physics 
may be stated as follows: Let $\Gamma (\phi, {\bfg}(\Lambda), \Lambda)$
 be  the effective action (or thermodynamical potential) obtained
 by  integration of the field variable with the energy scale down to
 $\Lambda$ from infinity or a very large cutoff $\Lambda _0$.
Here ${\bfg}(\Lambda)$ is a collection of the coupling constants including
 the wave-function renormalization constant defined at
 the energy scale at $\Lambda$. Then
the RG equation may be expressed as a simple fact that
 the effective action as a functional of the field variable $\phi$
 should be the same,  irrespective to how much the integration of the
 field variable is achieved, i.e.,
\beq
\Gamma (\phi, {\bfg}(\Lambda), \Lambda)=
\Gamma (\phi, {\bfg}(\Lambda '), \Lambda ').
\eeq
If we take the limit $\Lambda '\rightarrow \Lambda$, we have
\beq
\label{eq:1-wilson}
\frac{d\Gamma(\phi, {\bfg}(\Lambda), \Lambda)}{d\Lambda}=0,
\eeq
which is the Wilson RG equation\cite{wilson}, or the 
flow equation in the Wegner's terminology \cite{weg}; notice
 that Eq.({\ref{eq:1-wilson})
is rewritten as 
\beq
\frac{\d\Gamma}{\d\bfg}\cdot\frac{d\bfg}{d\Lambda}=
-\frac{\d\Gamma}{\d\Lambda}.
\eeq
If the number of the coupling constants is finite, the theory is called
 renormalizable. In this case, the functional space of the theory does not
 change in the flow given by the variation of $\Lambda$; one may say that
 the flow has an invariant manifold.

A notable aspect of the RG
  is that the RG equation gives a systematic tool for 
obtaining the infrared 
effective theories with fewer degrees of freedom than in 
the original Lagrangian relevant in the high-energy region. 
This is a kind of 
reduction of the dynamics. 
Finding  effective degrees of 
freedom and extract the reduced dynamics of 
the effective variables in fact 
have constituted  and still constitute the core 
of  various fields of theoretical physics.
 In QCD written in terms of the fields 
of quarks and gluons, the low energy effective theories in which 
 gluons are integrated out may be Nambu-Jona-Lasinio  type 
 lagrangians \cite{njl}; see for example \cite{hk94}.
 If the quarks are further integrated out the effective theories 
  are sigma 
models written solely with meson fields\cite{weinberg}: The flow equations,
 variants of the RG equation, may give a foundation on such sigma models
  as effective theories of QCD at low energies\cite{wetterich,morris,aoki};
  for a review, see \cite{rev_flow}: In terms of notions in the theory of 
  dynamical systems, 
the functional space represented by a sigma model may be
 an attractive submanifold of QCD.
 
Statistical physics may be said to be  a collection 
of  theories on how to 
reduce the dynamics of many-body systems to one with fewer variables,
 since the work of  Boltzmann\cite{boltzman}.
 Bogoliubov showed that BBGKY( Bogoliubov-Born-Green-Kirkwood-Yvon)
hierarchy can be reduced to Boltzmann equation with a single-particle
 distribution function for dilute gas systems.\cite{bogoliubov} 
 As indicated by Kuramoto\cite{kuramoto}, Bogoliubov seems to have claimed 
 that the dilute-gas dynamics has an attractive manifold spanned by 
 one-particle distribution function. 
 Boltzmann equation in turn can be further
  reduced to the hydrodynamic equation (Navier-Stokes equation)
   by a perturbation theory like 
  Chapman-Enskog 
  method\cite{chapman}.
  Recent development of the theories of pattern 
 formation with dissipative structures gives  a good example how to 
 reduce complicated  ordinary and partial differential equations to 
 simple equations with slow variables, such as Landau-Stuart equation,
 the time-dependent Ginzburg-Landau equation and so on.\cite{dissipative}

Some years ago,  it was shown  by an Illinois group\cite{cgo1,cgo2,cgo3}
 and Bricmont and Kupiainen\cite{bricmont}that the  RG equations
can be used  for a
 global and asymptotic analysis of
 ordinary and partial differential equations, hence giving a reduction 
 of evolution equations of some types.
A unique feature  of the Illinois group's
 method is to start with the naive 
 perturbative expansion and allow  secular terms to appear; the secular 
 terms correspond to the logarithmically divergent terms in QFT.
 Then introducing 
 an intermediate time $\tau$ and rewriting perturbative solutions 
 in terms of renormalization constants reminiscent of those 
 appearing in the  perturbative 
 renormalization theory in QFT, Gell-Mann-Low type RG equation is applied
 to obtain the evolution equations for the renormalization constants
  which are functions of $\tau$.
Finally, equating $\tau$ with the time $t$ appearing in the original 
perturbative solution, they obtained global solutions of differential 
equations.  Bricmont and Kupiainen \cite{bricmont} 
applied a scaling transformation 
(block transformation) to obtain asymptotic behavior of nonlinear diffusion
equations in a rigorous manner.

Subsequently, 
one of the present authors (T.K.) formulated  
 the Illinois group's
method in terms of the classical theory of envelopes \cite{kuni95,kuni97}:
 He indicated that the 
RG equation can be interpreted as the basic equation for constructing 
envelopes of a family of curves (or surfaces for partial differential 
equations). 
He also developed a short-cut prescription for the 
renormalization procedure 
 without introducing an intermediate time $\tau$ but utilizing 
an arbitrary initial time $t_0$. 
In  latest papers\cite{qm}, 
it was shown that the RG method
 can be well formulated as the method dealing with the initial values
 at arbitrary $t=t_0$; the initial values at $t=t_0$ are determined so that 
 the unperturbative solutions  which are valid only locally
  around $t=t_0$ are continued smoothly; this procedure is nothing but 
  to construct the envelope of the perturbative solutions.
He  emphasized that the RG method is a powerful tool for reduction 
of evolution equations and demonstrated it by applying the method to
obtain so-called the amplitude equations for systems of equations.
He also suggested that if 
 the phase equations describing slow motions in the system where 
  a continuous symmetry is broken, which are classical
  counter part of the Nambu-Goldstone modes in QFT, the RG method 
  should be able to derive the phase equations.

 The RG method developed by the Illinois group has been 
applied by many authors 
to quite a wide class of problems successfully. To mention
 some of them;
Graham\cite{graham} derived
 a rotationally invariant amplitude equation
  appearing  in the problem of pattern formation.
Some kinds of phase equations were also derived;
Oono \cite{phase} 
discussed the interface dynamics relevant in the spinodal decomposition,
Sasa derived a diffusion type phase equation,  and Maruo et al
\cite{maruo} derived  Kuramoto-Sivashinsky equation\cite{ks}.
The method was applied to analyze asymptotic behavior of
the non-linear equations appearing
 in cosmology\cite{cosmology1,cosmology2}. Boyanovsky and de Vega also
used the method to derive an anomalous transport coefficient
\cite{kinetic1} relevant in 
the  non-equilibrium states in the early universe and QGP (quark-gluon plasma).
Before that, the Boltzmann equation had been derived as an RG equation
\cite{kinetic2}. The RG method was also shown to be a powerful 
 tool to resum divergent 
 perturbation series appearing in problems of
 quantum mechanics\cite{frasca,qm}.
Tzenov  applied the method to obtain global solutions appearing accelerator
physics\cite{fermi}. Possible relation between renormalizability and
 integrability of Hamilton systems was discussed by Yamaguchi and 
Nambu\cite{hamilton}.
The method was proved to be applicable to discrete systems\cite{kunimatsu},
 too.
   
Although such extensive  applications have been made,
only few works are known which attempt to reveal the underlying reasons
 why the RG method works to some kinds of equations
 but not to others\cite{kuni95}. Nevertheless, it was indicated in 
 \cite{kuni97,kunimatsu}
 that the RG method by the Illinois group
 works when the unperturbed solutions are neutrally stable solutions
that are stationary (constant in time or stationary oscillation)
 hence do not decay nor blow up with time.
A decade ago, Kuramoto revealed in an excellent paper\cite{kuramoto}
 the universal underlying 
structure of all the existing perturbative methods for 
reduction of evolution equations;
 he noticed that when a reduction of evolution equation is possible,
 the unperturbed equation admits neutrally stable solutions, and 
 succeeded in describing the reduction of dynamics in geometrical terms, i.e.,
 attractive manifolds or invariant manifolds\cite{holmes}.
Although his actual 
presentation of the theory was based on the reductive perturbation theory
and involves some ansatz on forms of the solutions,
he emphasized that the universal structure which he revealed should not be
dependent on the perturbation methods one employs.  
In the present paper, focusing on the aspect of the RG method as a powerful 
tool for reducing evolution equations, we shall 
 present  a comprehensive 
formulation of the perturbative RG method in terms of the notion of
invariant manifolds, guided by this Kuramoto's work:
 One will see that his
ansatz are {\em derived} naturally in the RG method.
It may mean that his
formulation of reduction of evolution equations,
 which is actually a natural
  extension of the asymptotic method by Krylov,
Bogoliubov and Mitropolski for non-linear oscillators\cite{kbm},
is an RG theory although the term RG is not used\cite{kinetic2}.
 
We  start  with adapting the exact RG equations
of Wilson type  (flow equations) in quantum field theory \cite{wilson,weg}
 to differential equations (evolution equations); one may recognize
  that  the RG method described in terms of envelopes is best formulated
in the framework of Wilson RG.
Then  confining ourselves to cases where a perturbative treatment 
is possible, we shall show that an invariant manifold exists
 when the RG method works, and that 
the RG method  is a method to construct the invariant manifold and the 
reduced dynamics on it in a mechanical way. In this method,
the initial values of the solution at an arbitrary time $t=t_0$ 
 in the successive perturbation orders 
are determined so that the initial value  make an invariant manifold,
thereby the condition to set $t_0=t$ will be naturally emerged.
We emphasize that the present formulation 
give a nice foundation for the prescriptions adopted 
in \cite{kuni95,kuni97,qm}.

The following should be  mentioned here: 
(1) The relevance of Wilson type RG equation to 
their method was noted by Chen et al\cite{cgo2} and Pashkov and 
Oono\cite{kinetic2}, although their formulation was totally based on the 
Gell-Mann-Low type perturbative RG method.  
(2) Shirkov \cite{shirkov} 
clarified that the RG equation concerns with the initial values and 
 emphasized the Lie group structure of the RG method\cite{lie}; he
extracted the notion of {\em functional self-similarity} 
 (FSS) as the essence of the exact RG. He claims that the
 Wilson RG is an approximation to the Bogoliubov
RG which is exact\cite{shirkov}.
 
Once the underlying structure of the reduction of dynamics given by the RG 
method has become clear, one will recognize that the method  
could be applied to problems which have not been treated in the RG method 
 so far although other methods  are applied to them.
 One will also see that the RG method is simpler to 
 apply to them than the previous methods. As such a problem, we take the
 problem to extract the interface interactions\cite{ko,cp,eo,FH},
 which are typical examples to be treated by the method of phase equations.  

In \S 2, we formulate the RG method  as a method of reduction of dynamics,
 starting from the non-perturbative flow equation (RG equation).
 Then on the basis of the perturbation theory, we show, guided by the
presentation given in \cite{kuramoto}, the way  how the invariant 
 manifold which is supposed to exist to the evolution equation under
 consideration can be constructed in our method. 
 We shall remark that
  the existence of an invariant manifold corresponds to the notion of 
  the renormalizability in quantum field theory.  
 In \S 3, some simple but typical examples including 
 the Takens equation \cite{takens} are worked out 
 to demonstrate how the RG method construct invariant manifolds and
  give the reduced dynamics on them.
 In \S 4, we show how to deal with generic systems which involve  a 
 linear operator $A$ having zero-eigenvalue where  $A$ may or may not 
 be diagonalizable; when $A$ is not diagonalizable, the eigenvalues are
degenerate and $A$ is equivalent with a matrix having a Jordan cell
$\pmatrix{0\  1\cr 0\ 0}$.
 As examples with a Jordan cell, we shall show that
 the RG equation gives the 
normal forms \cite{wiggins}
 of the two-dimensional equations including the Takens and 
 the Bogdanov equation \cite{bogdanov}, and deal with  
an extended version of Takens equation with three-degrees of freedom.
 In \S 5, we apply the method to some problems such as 
  the unstable motion
in the Lotka-Volterra system and the Hopf bifurcation in the 
Brusselator\cite{brussel}.
Although the examples treated in \S 2-5 are simple ordinary differential
 equations, we believe that these examples should be instructive also 
 for experts of the RG's or flow equations 
 in quantum field theory and/or statistical physics.
 In \S 6, we also apply the method to extract 
the interface dynamics of a kink-anti-kink 
interaction in the TDGL equation in one-dimension and the soliton-soliton
 interaction in the KdV equation.
 The final section is devoted to a brief summary and concluding remarks.
In Appendix A, we summarize rules in a scheme of an operator method
for constructing special solutions 
suitable for the RG method.
 In Appendix B, we present an elementary method different from the 
RG method  to 
derive the 
 approximate solution for the double-well potential discussed in 
subsection 3.3.
In Appendix C, we show that the period of the Lotka-Volterra system 
\cite{lv} around 
the non-trivial fixed point obtained previously\cite{kuni97} in the RG method 
coincides with that 
extracted by Frame \cite{frame} in a quite different way.

\newpage
\section{Reduction of evolution equations with  the RG method}
\renewcommand{\theequation}{\thesection.\arabic{equation}}
\setcounter{equation}{0}

In this section, we shall formulate the RG 
method in terms of the notion of  invariant  manifold.

\subsection{The RG equation as the flow equation for initial values}

\subsubsection{Non-perturbative RG equation}

Let us take  the following $n$-dimensional dynamical system;
\beq
\label{sect2;1}
\frac{d \bfX}{dt} = \bfF(\bfX, t),
\eeq
where $n$ may be infinity.
We remark that 
 $\bfF$ may depend on $t$ explicitly.
We suppose that the 
equation is solved up to an arbitrary time 
 $t=t_0$ from an initial time, say, at $t=0$,
  to give $\bfX(t)=\bfW(t)$, and then we are
  trying to solve the equation with the initial condition 
 at $t=\forall t_0$,
\beq
\label{init0}
\bfX(t=t_0)=\bfW(t_0),
\eeq
with $\bfW(t_0)$ being unspecified yet.
In fact,  $\bfW(t)$ as a function of $t$ 
is a solution of (\ref{sect2;1}) by definition.
The solution may be written as
$\bfX(t; t_0, \bfW(t_0))$. 
We stress that the solution needs not be given by a perturbation method;
if the solution is given non-perturbatively, the resultant equations
remain non-perturbative ones.

Now, making use of  $\bfX(t; t_0, \bfW(t_0))$ thus obtained, we determine
$\bfW(t_0)$ based on a simple fact of differential equations.
When the  initial point is shifted to $t_0'$, the resultant solution 
  should be the same, i.e.,
\beq
\bfX(t; t_0, \bfW(t_0))=\bfX(t; t_0', \bfW(t_0')).
\eeq  
Taking the limit $t_0' \ \rightarrow \ t_0$, we have\cite{qm} 
\beq
\label{rg00}
\frac{d \bfX}{dt_0} =\frac{\d \bfX}{\d t_0}+
\frac{\d \bfX}{\d \bfW}\frac{d \bfW}{dt_0}={\bf 0}.
\eeq
This equation gives the evolution equation or
the flow equation of the initial value $\bfW(t_0)$.
This has the same form as that of the renormalization group (RG) equation
 in quantum field theory, hence the name of the RG method. 
We emphasize that the equation (\ref{rg00}) is exact; we have not used
any argument based on perturbation theories.
This equation corresponds to the non-perturbative RG equations 
(flow equations) by Wilson\cite{wilson}, Wegner-Houghton\cite{weg}
 and so on in quantum field
 theory and statistical physics\cite{rev_flow}:
The reader should have recognized that $t_0$ corresponds to 
the logarithm of the energy scale
to which the integration is performed in the quantum field theory.
The quantities corresponding to the coupling constants will be found to 
be the integration constants.  One will also recognize that the existence 
of an invariant manifold of a dynamical system corresponds to the 
renormalizability. 
We also notice that one needs
not to equate $t_0$ with $t$ at this stage.

\subsubsection{Perturbative RG equation}

So far, only the perturbative expansion
 method is available  to make $\bfX(t; t_0, \bfW(t_0))$.
 In this case, $\bfX (t;t_0, \bfW(t_0))$ and $\bfX (t;t'_0, \bfW(t'_0))$ 
 may  be valid 
only for $t\sim t_0$ and $t\sim t'_0$. This condition is naturally 
satisfied if we restrict that $t_0<t<t'_0$ (or $t'_0<t<t_0$) because
the limit $t'_0 \rightarrow t_0$ is to be taken.
Thus when a perturbative expansion is used for constructing 
$\bfX (t; t_0, \bfW (t_0))$,
we demand a more restrictive RG equation as given by\cite{qm}
\beq 
\label{rg0}
\frac{d \bfX}{dt_0}\biggl\vert _{t_0=t} = \frac{\d \bfX}{\d t_0}
\biggl\vert _{t_0=t} +\frac{\d \bfX}{\d \bfW}\frac{d \bfW}{dt_0}
\biggl\vert _{t_0=t} ={\bf 0}.
\eeq
Notice that the demand to set $t_0=t$ has naturally emerged.

We remark that a geometrical interpretation of this equation has been 
given on the basis  of the classical theory of envelopes
\cite{kuni95,kuni97,qm}:
 When $t_0$ is varied, $\bfX(t; t_0, \bfW(t_0))$  
 gives a family of curves with $t_0$ being a parameter characterizing
curves. Then Eq.(\ref{rg0}) is the condition  to construct the envelope of 
the family of curves which are valid only locally around $t\sim t_0$.
The envelope is given by $\bfX (t;t_0=t)=\bfW (t)$, i.e, the initial value.
It is noteworthy that $\bfW(t)$ satisfies the original equation 
(\ref{sect2;1}) in a global domain up to the order with which 
$\bfX(t; t_0)$ satisfies around $t\sim t_0$\cite{kuni95,kuni97}. 
In fact, one can see that 
\beq
\frac{d\bfW}{d t}&=&\frac{\d \bfX(t;t_0)}{\d t}\biggl\vert_{t_0=t}
 +\frac{\d \bfX(t;t_0)}{\d t_0}\biggl\vert_{t_0=t},\nonumber \\ 
 \ &=& \frac{\d \bfX(t;t_0)}{\d t}\biggl\vert_{t_0=t},
 \eeq
 on account of (\ref{rg0}).

\subsection{Invariant manifolds and renormalizability}

In this section, we follow \cite{kuramoto} for notations.
If the theory has an invariant manifold M with
 the dimension less than $n$, we may have supposed that  the initial point 
 is on the manifold.
Let the invariant manifold M is represented by the coordinate $\bfs $.
 The reduced dynamics of Eq.(\ref{sect2;1}) on M
may be  given in terms of a vector field ${\bfG}$ by
\beq
\frac{d \bfs }{dt} = {\bfG}(\bfs ),
\eeq
and the manifold M is represented by
\beq
\bfX=\bfR(\bfs ).
\eeq
Our task is to obtain the vector field ${\bf G}$ and the representation
 of the manifold $\bfR$ in a perturbation method.
We consider a situation where the vector field $\bfF$ 
is composed of an unperturbed part $\bfF_0$ and the perturbative one
 $\bfP $, i.e.,
\beq
\label{sect2;eq}
\bfF= \bfF_0(\bfX) + \eps \cdot \bfP (\bfX, t).
\eeq
Here notice that $\bfF_0$ has no explicit $t$-dependence, while
 $\bfP (\bfX, t)$ does.
 We assume that the unperturbed problem is solved and an attractive 
invariant manifold M$_0$ is easily found.  

Now  we try to solve Eqs.(\ref{sect2;1}) and (\ref{sect2;eq}) by a 
perturbation theory  with the initial condition 
\beq
\bfX(t_0)= \bfW(t_0),
\eeq
at $t=t_0$.  The decisive point of our method is to assume that
\beq
 \bfW(t_0)=\bfR(\bfs(t_0)),
\eeq
that is, the initial point is supposed to be on the invariant manifold M 
to be determined.
 Now we apply the perturbation theory, expanding  
\beq
\bfX(t; t_0, \bfW(t_0))=
 \bfX_0 + \sum \eps ^n \bfX_n(t; t_0, \bfW(t_0)).
\eeq
Here we have made it explicit that $\bfX$ is dependent on the initial
 condition.  
 We should also expand the initial value,
\beq
\bfW(t_0)= \bfW_0(t_0) + \brho (t_0),
\eeq
with
\beq
 \brho (t_0)=\sum _{n=1}^{\infty}\eps^n \bfW_n(t_0).
\eeq

Now the unperturbed equation reads
\beq
\frac{d \bfX_0}{dt} = \bfF_0(\bfX_0).
\eeq
As promised, we suppose that an attractive manifold is found for this
equation as
\beq
\bfX_0(t)=\bfR(\bfs (t; \bfC)),
\eeq
where $\bfC$ is the integral constant vector with dim$\bfC=m$ and
 may depend on $t_0$.
 
 Here comes an important point of our method; we identify that 
\beq
\bfs  (t_0)=\bfC(t_0),
\eeq
which gives a natural parameterization of the manifold M$_0$.  This is 
 a simple but a significant observation; notice that we need not to 
 give any ansatz to the representation of the manifold because we only
 have to solve the unperturbed equation and the integral constants are 
 trivially obtained.
 
The deformation of the manifold ${\bfrho}$ will be 
determined perturbatively on the two principles, 
i.e.\cite{kuramoto},
\begin{enumerate}
\item the function ${\bfrho}$ should be independent of $\bfW_0$, and
\item the resultant dynamics 
should be as simple as possible because we are interested to reduce the
dynamics to a simpler one. 
\end{enumerate}

The choice of the ${\bfrho}$ is intimately
related to that of the forms of the perturbative solutions.
For example,
 the first order equation reads
\beq
\frac{d \bfX_1}{dt} = \bfF'_0(\bfX_0)\bfX_1 +\bfP (\bfX_0).
\eeq
The solution to this inhomogeneous equation is composed of a sum of 
the general solution of the homogeneous equation and the special 
solution of the inhomogeneous equation.  If the unperturbed solution 
$\bfX_0(t)$ has a part of neutrally stable solution, there appear
secular terms in the special solution as well as 
genuinely independent functions. It turns out that 
the secular terms can be utilized to renormalize out 
the homogeneous solutions 
 at $t=t_0$\cite{kuni95,kuni97}; this is a kind of renormalization 
conditions.
  Then the shift of
the initial value is now determined as 
\beq
\bfW_1(t_0)= \bfX_1(t=t_0).
\eeq
Notice that the initial value is determined after solving the equation, 
hence
 the functional form of it as a function of $t_0$ is explicitly given.
Of course, one can make $\bfW_1(t_0)= 0$ by further adding unperturbed 
solutions; this prescription, however, will generally 
give a more complicated dynamics.  It means that the choice of ${\bfrho}$
 has ambiguities, but
the demand to obtain the simplest dynamics give it uniquely.
In \S 3 and 4, we shall give a mathematical and mechanical procedure to select
the initial values in accordance with the above rules using some 
 examples.

We can repeat the procedure to any order of the perturbation.
We remark that by this procedure the modification of the initial
value $\bfrho(t_0)$ and hence the total initial value $\bfW(t_0)$ are
given solely in terms  of $\bfC(t_0)$;
\beq
\bfW(t_0)=\bfW_0[\bfC]+\bfrho[\bfC].
\eeq

Now the dynamics of $\bfC(t)$ is determined by the RG equation 
Eq.(\ref{rg0});
\beq
\label{rg1}
\frac{d \bfX}{dt_0}\biggl\vert_{t_0=t} =
\frac{\partial \bfX}{\partial t_0}\biggl\vert_{t_0=t}
+\frac{\partial \bfX}{\partial \bfC}\cdot \frac{d \bfC}{dt_0}
\biggl\vert_{t_0=t}
=0,
\eeq
which is an evolution equation of $\bfC(t)$.
Then the manifold M  (more precisely, the trajectory on it) is represented 
as
\beq
\label{eq:sect2-final}
\bfX =\bfW(t)=\bfW_0[\bfC(t)]+\bfrho[\bfC].
\eeq
Eq.'s (\ref{rg1}) and (\ref{eq:sect2-final}) are essentially the
 basic postulates in  Kuramoto's theory 
of reduction of evolution equations\cite{kuramoto}.
Thus we have derived the Kuramoto's
basic equations in the RG method; in other words,
Kuramoto's theory is actually the
RG theory for reduction of evolution equations.
Eq.(\ref{eq:sect2-final})
 means that the dynamics is renormalized to that for $\bfC$.
One can now see  a correspondence between the 
renormalizability of the theory in quantum field theory\cite{weinberg} 
and the existence of a finite dimensional 
invariant manifold in the theory of dynamical systems:
$\bfC=(C_1, C_2, \dots, C_m)$
 correspond to the collection of renormalizable
coupling constants and $\bfrho$ to unrenormalizable operators;
one may say that 
the fact that $\bfrho$ can be represented solely  with $\bfC$ is
 analogue of the Polchinski 
theorem\cite{pol,weinberg} in quantum field theory.
We remark also 
 that  Eq.(\ref{eq:sect2-final})
justifies the slaving principle by Haken\cite{haken}.

A comment is in order; When the unperturbed system is given by a linear
 operator with a Jordan cell, there will be a slight modification of the 
 above scenario due to a 
technical complexity;
 see an example in the next section and the general argument given in \S 4.
\setcounter{equation}{0}
\section{Simple examples}

In this section, we  consider four simple equations
 to show  our
 formulation of the RG method at work 
as a tool for reduction of evolution equations.
We shall show how the initial values at $t=\forall t_0$ in higher orders
 are determined
 by the two principles that terms proportional to the 
 unperturbed solution are 
suppressed and that possible fast motions disappear.
 The resultant forms of the unperturbed solutions composed 
of the secular terms  which vanish at $t=t_0$ and
 the functions independent of the unperturbed solution.
The final forms turn out to be
  the same as those given in the scheme 
 adopted in \cite{kuni95,kuni97,qm}, where unstable manifolds and cases
 with a Jordan cell were not dealt with  though.
The resultant initial value represent the invariant manifold, and the
 RG equation gives the slow dynamics on the manifold.
 
\subsection{A simple model with a reduction}

We  first consider the following simplest equation 
with a reduction\cite{kuramoto};
\beq
dx/dt=\eps f(x, y), \ \ \ dy/dt= -y +g(x).
\eeq
Writing $\bfu (x, y)=\ ^t(x, y)$  and we expanding $\bfu $ as
$\bfu =\bfu _0 + \eps \bfu _1+ \cdots,$
with $\bfu _n(x, y)=\ ^t(x_n, y_n)$\ $(n=0, 1, ...)$. We solve the
 equation with the initial value $\bfW(t_0)$ at $t=t_0$.
We suppose that $\bfW(t_0)$ is on an attractive manifold M.

The unperturbed equation reads
\beq
\dot{x}_0=0,\quad \dot{y}_0=-y_0+g(x_0),
\eeq
the  solution to which is readily obtained as
$x_0(t)={\rm const.}=C_0,\ \quad y_0(t)=g(C_0)+ C_1{\rm e}^{-t}$,
with $C_i\ (i=0, 1)$ being the integral constants;
we make it explicit that $C_i$ may depend on the 
initial time $t_0$ as $C_i=C_i(t_0)$.
Since we are interested in the asymptotic behavior as 
$t\rightarrow \infty$, we take the stationary solution 
putting $C_1=0$;
\beq
x_0(t)={\rm const.}=C_0,\ \quad y_0(t)=g(C_0),
\eeq
accordingly, 
\beq
\bfW_0(t_0)=\ ^t(C_0, g(C_0)),
\eeq
which suggests that the unperturbed invariant manifold M$_0$ is given by 
\beq
 y=g(x),
\eeq
although the time dependence $x(t)$ is not yet known.

The first order equation reads
\beq
(\d_t - A){x_1\choose y_1}={f(x_0, y_0)\choose 0},
\eeq
with
\beq
A=\pmatrix{0\quad\ \quad 0\cr
           g'(x_0)\ -1}.
\eeq
We notice that 
\beq
A\bfU_1=0,\quad A\bfU_2=(-1)\bfU_2 ;\quad
 \bfU_1={1\choose g'},\quad \bfU_2={0\choose 1}.
\eeq
Then the  special solution with the initial value
 $\bfW_1(t_0)$  is obtained as follows:
\beq
{x_1\choose y_1}&=&\e^{(t-t_0)A}\bfW_1(t_0)+\int_{t_0}^tds\e^{(t-s)A}
                 {f(x_0, y_0)\choose 0},\nonumber \\ 
                &=& \e^{(t-t_0)A}[\bfW_1(t_0)+fg'\bfU_2]+
          \{(t-t_0)f\bfU_1-
                    fg'\bfU_2\}.
\eeq
Here we have used the relation that $\ ^t(1, 0)=\bfU_1-g'\bfU_2$.
For the solution to describe a slow motion, 
 the first term should vanish; thus we are naturally led 
to the choice of the initial value as 
\beq
\bfW_1(t_0)= -fg'\bfU_2,
\eeq
to kill the fast  motion.
 Thus we have also
\beq
\label{eq:ex-1}
x_1(t;t_0)=f(C_0,g(C_0))(t-t_0),
\quad y_1(t;t_0)=f(C_0,g(C_0))g'(C_0)\{(t-t_0)-1\}.
\eeq
We notice that the solution and the initial value
 satisfy the rules given in the preceding 
subsection; $\bfW_1(t_0)\simeq\bfrho(t_0)$ is independent of $\bfW_0(t_0)$.

Up to this order,
\beq
\label{pertresult:simple}
\bfu (t;t_0)=\pmatrix{x(t;t_0)\cr
                      y(t;t_0)}
            =\pmatrix{C_0(t_0)+\eps f(C_0, g(C_0))(t-t_0)\cr 
                      g(C_0)- \eps g'(C_0)f(C_0, g(C_0))(1+ t_0-t)
                      },
\eeq
with
\beq
\bfW (t_0)= {x(t_0)\choose y(t_0)}
\simeq \bfW_0(t_0)+\bfrho(t_0)=
 {C_0\choose{g(C_0)-\eps g'(C_0)f(C_0, g(C_0))}}.
\eeq

Now the RG equation Eq.(\ref{rg0}) applied to 
Eq.(\ref{pertresult:simple}) gives 
the evolution equation for $C_0(t)$;
\beq
\label{rg2}
\frac{d C_0}{d t}= \eps f(C_0, g(C_0)).
\eeq
Since $x(t)=C_0(t)$, Eq.(\ref{rg2}) gives the reduced dynamics, and
the slow manifold is given by $\bfu(t)=\bfW(t)$, or in terms of the
components,
\beq
x(t)=C_0(t), \ \ y(t)= g(C_0)-\eps g'(C_0)f(C_0, g(C_0))).
\eeq
One sees that the attractive manifold is given by 
\beq
y(x)=g(x)- \eps g'(x) f(x, g(x)),
\eeq
and the original two-dimensional evolution equation 
 has been reduced to the one-dimensional equation Eq.(\ref{rg2}).

In short, to obtain the invariant manifold and the slow dynamics 
 on it, we have started with a neutrally stable solution with
 an integral constant, which constitutes a natural representation of the
 invariant manifold. The constant moves slowly by the 
 perturbation and the whole dynamics is given through this slow
 variable. The evolution equation of the slow dynamics is given 
 by the RG equation and the trajectory on the invariant manifold is
  given as the initial value in the RG method uniquely.
The initial value in the higher order 
 has been determined so that the fast motion disappears.

Generic systems which have a linear matrix having zero eigenvalue
will be extensively analyzed in \S 4.

\subsection{When the unperturbed solution is oscillatory}

 In the present subsection,
we shall examine a case  where the unperturbed solution has  another type of
 neutral stability, i.e., is  oscillatory .
We shall treat a simplest equation of a damped oscillator 
with the  
geometrical terms putting an emphasis on the fact that the RG method 
concerns with the initial value. 
 In a recent monograph, Nishiura\cite{nishiura}
 used this example to indicate  that a careful identification of the
initial values in the higher order terms is needed to construct 
the proper perturbative solutions for obtaining a meaningful 
result in the RG method, though he failed to  give general principles for 
``a careful identification''. 
One will see that the general principles given in the foreword
 in this section determine the initial values uniquely, hence Nishiura's
 concern is resolved.
One will also  see that 
 the initial values thus obtained  are 
of   the same forms as 
 obtained  in \cite{kuni95}.

The equation  we deal with is
\beq
\label{eqdump}
\ddot{x}+\eps \dot{x} +x=0,
\eeq
 where $0<\eps<1$. 
This system does not exhibit
a decrease of the degrees of freedom, nevertheless the dynamics is 
reduced to a set of
 simpler equations for the amplitude and the phase, separately
 by the RG equation.

With the definition $\bfu =\ ^t(x, y),\ y=\dot{x}$, 
Eq.(\ref{eqdump})is converted to 
\beq
(\d_t-A)\bfu=-\eps{0\choose y},
\eeq
with
\beq
A=\pmatrix{0\ \ \ 1\cr
           -1\ 0}.
\eeq

We expand the dependent variable and the initial value in Taylor series as
$\bfu (t)=\bfu_0+\eps \bfu_1 +\eps^2\bfu_2 +\cdots$ and 
$\bfW (t_0)=\bfW_0+\eps \bfW_1 +\eps^2\bfW_2 +\cdots,$
with $\bfu_i=\ ^t(x_i, y_i)$.
The lowest order solution reads with the (complex) integral constant
$C(t_0)$;
\beq
\bfu _0(t;t_0)=C\e ^{it}\bfU_{+}+C^{*}\e ^{-it}\bfU_{-}\equiv
     {x_0(t; t_0)\choose y_0(t; t_0)},
\eeq
where $\bfU_{\pm}=\ ^t(1, \pm i)$; $A\bfU_{\pm}=\pm i\bfU_{\pm}$.
This is a neutrally stable solution.
The initial value in  this order  reads,
\beq
\bfW _0(t_0)=z(t_0)\bfU_{+} + {\rm c.c.},
\eeq
with 
\beq
z(t_0)=C(t_0)\e^{it_0}.
\eeq
 Here
 c.c. denotes the complex conjugate. 

The first order equation reads
\beq
(\d_t-A)\bfu_1=-y_0(t){1\over {2i}}(\bfU_{+}-\bfU_{-}),
\eeq
with the initial condition $\bfu_1(t_0, t_0)=\bfW_1(t_0)$ which is 
not yet known but to be determined.
The equation is readily solved as follows; 
\beq
\label{dump;sol-1}
\bfu_1(t;t_0)&=&\e^{(t-t_0)A}\bfW_1(t_0)
             -{1\over {2i}}
       \int_{t_0}^tds\e^{(t-s)A}y_0(s)(\bfU_{+}-\bfU_{-}),
       \nonumber \\
        &=& \e^{(t-t_0)A}\left[\bfW_1(t_0)-{1\over {2i}}\{C^*\e^{-it_0}
        \bfU_{+}
        -C\e^{it_0}\bfU_{-}\}\right]\nonumber \\
       & &
        -{1\over 2}
        \left[\{(t-t_0)C\e^{it}\bfU_{+}-{1\over{2i}}C\e^{it}\bfU_{-}\}+
        {\rm c.c.}\right],.
\eeq
which leads to the natural choice of the initial value
\beq
\bfW_1(t_0)={1\over {2i}}\{z^{*}(t_0)\bfU_{+}
        -z(t_0)\bfU_{-}\}=\bfrho_1[z, z^*],
\eeq
because otherwise the first term of (\ref{dump;sol-1}) gives rise to
terms which could be renormalized away into the unperturbed solution
 with a redefinition of $C$. We emphasize that a renormalization procedure
  enters here.
Hence
\beq
\bfu_1(t;t_0)=-{1\over 2}\{(t-t_0)C(t_0)\e^{it}\bfU_{+}-{1\over{2i}}C(t_0)
    \e^{it}\bfU_{-}\}+
        {\rm c.c.}\equiv{x_1(t;t_0)\choose y_1(t;t_0)}.
\eeq
        
Similarly, the second order solution is given by
\beq
\bfu_2(t;t_0)&=&\e^{(t-t_0)A}\bfW_2(t_0)
       -{1\over {2i}}\int_{t_0}^tds\e^{(t-s)A}y_1(s)(\bfU_{+}-\bfU_{-}),
       \nonumber \\
       &=&\e^{(t-t_0)A}\biggl[\bfW_2(t_0)-\{-\frac{C}{16}\e^{it_0}\bfU_{-}
        + {\rm c.c.}\}\biggl]\nonumber \\
       & & + \frac{iC}{8}[\{i(t-t_0)^2+(t-t_0)\}\bfU_{+}
             +\{t-t_0 +i/2\}\bfU_{-}]\e^{it} + {\rm c.c.}.
\eeq
Thus we are led to the choice of the initial value
\beq
\bfW_2(t_0)=-\frac{1}{16}z(t_0)\bfU_{-}
        + {\rm c.c.}=\bfrho_2[z,z^*],
\eeq
because of the same reason as in the first-order case.
Hence
\beq
\bfu_2(t;t_0)= \frac{iC}{8}[\{i(t-t_0)^2+(t-t_0)\}\bfU_{+}
            + \{t-t_0 +i/2\}\bfU_{-}]\e^{it} + {\rm c.c.}.
\eeq

Collecting all the terms up to the second order, we have 
\beq
\bfu(t;t_0)&=&C\e^{it}\bfU_{+}-{{\eps C}\over 2}\{(t-t_0)\bfU_{+}
            +{i \over 2}\bfU_{-}\}\e^{it},\nonumber \\
 \ \ \ & & +\frac{i\eps^2C}{8}[\{i(t-t_0)^2+(t-t_0)\}\bfU_{+}
             +\{t-t_0 +i/2\}\bfU_{-}]\e^{it} + {\rm c.c.},
\eeq
with  the initial value 
\beq
\bfW(t_0)=z(t_0)\{\bfU_{+}-i\frac{\eps}{4}\bfU_{-}
-\frac{\eps^2}{16}\bfU_{-}\} +{\rm c.c.}.
\eeq

The RG equation Eq.(\ref{rg1})  gives
\beq
\dot{C}+(\eps/2+ i\eps^2/8)C=0.
\eeq
Parameterizing $C$ as $C=(A/2)\cdot{\rm exp}(i\theta)$,
we have the equations governing the amplitude and the phase, respectively,
\beq
\label{reddump}
\dot{A}+\eps/2\cdot A=0, \ \ \ \dot{\theta}=-\eps^2/8,
\eeq
which yields
$A(t)=A_0\e^{-\eps/2\cdot t}$ and  $\theta(t)=-\eps^2t/8+\theta_0$
with $A_0$ and $\theta_0$ being constants. 
Since $\bfu (t)=\bfW (t)$, we have the final solution to the damped 
oscillator as
\beq
x(t)=A_0\e^{-\eps/2\cdot t}\{(1-\frac{\eps^2}{16})\cos (\omega t+\theta_0)
+\frac{\eps}{4}\sin (\omega t+\theta_0)\},
\eeq
with $\omega=1-\eps^2/8$ being the angular velocity.
The above expression is slightly 
different from that given in \cite{kuni95} where the equation is treated 
as a scalar equation;
 a redefinition of the constants $A_0$ and $\theta_0$ transforms
 the solutions to each other in this order.
We see that although the number of the dimension of the equation is 
not changed, the dynamics is reduced to simpler equations
(\ref{reddump}) for the amplitude and the phase.
The initial values in the higher order equations
 have been determined so that terms proportional to the unperturbed solution
 do not appear; such higher order terms have been "renormalized away'' by a
 redefinition of the integral constant in the unperturbed solution.

A comment is in order:
The above solution could be  more
efficiently obtained by using the operator method (implicitly) adopted 
in \cite{kuni95,kuni97} and fully accounted 
 in Appendix A of the present article.
For instance, the first order solution in the operator method reads
\beq
\bfu_1(t;t_0)=-{1\over 2}(\d_t-A)^{-1}(C\e^{it}-C^*\e^{-it})
(\bfU_{+}-\bfU_{-}),
\eeq
which is readily evaluated with the use of (A. 9) and (A. 10).
The second order solution is also readily obtained with the use of the 
formulae given in Appendix A.

\subsection{Unstable motion in the double-well potential}

Next, we show that unstable manifolds are also constructed by the present 
method, using the double-well potential in mechanics;
\beq
\ddot{x}=x-\eps x^3.
\eeq
Let us obtain the unstable motion around the origin by applying the 
RG method.  This example is a peculiar case where the unperturbed
solution is not neutrally stable but composed of a blowing and decaying
 function with the same exponent. We shall treat another example of this 
type in \S 5.2.

Putting $\dot{x}=y$ and defining $\bfu=\, ^t(x, y)$, we have a system of 
equation
\beq
\left({d\over{dt}}-A\right)\bfu=\eps{0\choose {-x^3}}.
\eeq
We solve the equation around an arbitrary $t=t_0$ with the initial value
$\bfW(t_0)$.
The solution is written as $\bfu=\bfu(t;t_0,\bfW(t_0))$.
We expand $\bfu=\bfu_0+\eps\bfu_1+\cdots.$  Accordingly the initial value
$\bfW(t_0)$ which is to be determined self-consistently is also expanded as
$\bfW=\bfW_0+\eps\bfW_1+\cdots.$

The lowest order solution reads
\beq
\bfu_0(t;t_0)=C_{+}(t_0)\e^t\bfU_{+}+C_{-}(t_0)\e^{-t}\bfU_{-},
\eeq
where we have suppressed the $\bfW$-dependence of $\bfu_0$ and 
$\bfU_{\pm}=\, ^t(1, \pm 1)$ are the eigenvectors of $A$ belonging 
to the eigenvalues $\pm 1$, respectively.
The initial value $\bfW_0(t_0)$ accordingly reads
\beq
\bfW_0(t_0)=
\bfu_0(t_0;t_0)=C_{+}(t_0)\e^{t_0}\bfU_{+}+C_{-}(t_0)\e^{-t_0}\bfU_{-},
\eeq
which will imply that the trajectory is in a hyperbolic curve
\beq
{\rm M}_0=\{\bfu=(x, y)\vert (x+y)(x-y)={\rm const}\}
\eeq
with the asymptotic lines $y=\pm x$.

The first order equation reads
\beq
\left({d\over{dt}}-A\right)\bfu_1=-{1\over 2}(C_{+}(t_0)\e^t+C_{-}(t_0)
\e^{-t})^3
    (\bfU_{+}-\bfU_{-}),
\eeq
which is solved with the initial value $\bfW_1(t_0)$ formally as
\beq
\label{dw;sol-1}
\bfu_1(t;t_0)& = & \e^{(t-t_0)A}\bfW_1(t_0) \nonumber \\  
   & & -{1\over 2}\int_{t_0}^tds\e^{(t-s)A}(C_{+}(t_0)\e^s+C_{-}(t_0)
    \e^{-s})^3 (\bfU_{+}-\bfU_{-}),\nonumber \\ 
   &=& \e^{(t-t_0)A}[\bfW_1(t_0)
    +{1\over 2}(f_{+}(t_0;t_0)\bfU_{+}+f_{-}(t_0;t_0)\bfU_{-})]\nonumber \\ 
   & & -{1\over 2}(f_{+}(t;t_0)\bfU_{+}+f_{-}(t;t_0)\bfU_{-}),
\eeq
with 
\beq
f_{+}(t;t_0)&=&3C^2_{+}C_{-}(t-t_0)\e^t-{3\over 2}C_{+}C^2_{-}\e^{-t}
             +{1\over 2}C^3_{+}\e^{3t}-{1\over 4}C^3_{-}\e^{-3t},\\ 
f_{-}(t;t_0)&=&-3C_{+}C^2_{-}(t-t_0)\e^{-t}-{3\over 2}C^2_{+}C_{-}\e^{t}
             +{1\over 2}C^3_{-}\e^{-3t}-{1\over 4}C^3_{+}\e^{3t}.
\eeq
Thus the first-order initial value $\bfW_1(t_0)$ is chosen to be
\beq
\bfW_1(t_0)=-{1\over 2}(f_{+}(t_0;t_0)\bfU_{+}+f_{-}(t_0;t_0)\bfU_{-})
          ,
\eeq
because otherwise the first term of (\ref{dw;sol-1}) gives rise to 
 terms proportional to the unperturbed solution which should be 
"renormalized away'' with the redefinition of $C_{\pm}$.
 We notice that $\bfW_1(t_0)$ is a function of
 $A_{\pm}(t_0)=C_{\pm}(t_0)\e^{\pm t_0}$;
\beq
\bfW_1(t_0)=\bfrho_1[A_{+}, A_{-}].
\eeq
 
Applying the RG equation to $\bfu(t;t_0)=\bfu_0(t;t_0)+\eps \bfu_1(t;t_0)$,
one obtains
\beq
{{d\bfu}\over{dt_0}}\biggl \vert _{t_0=t}=\dot{C}_{+}\e^t\bfU_{+}+
 \dot{C}_{-}\e^{-t}\bfU_{-}+3{\eps\over 2}\{C^2_{+}C_{-}\e^t\bfU_{+}
 -C_{+}C^2_{-}\e^{-t}\bfU_{+}\}=0,
\eeq
which leads to
\beq
\dot{C}_{+}=-{{3\eps}\over 2}C^2_{+}C_{-},\quad 
\dot{C}_{-}={{3\eps}\over 2}C_{+}C^2_{-}.
\eeq
Noting that $C_{+}(t)C_{-}(t)={\rm const}\equiv c_{+}c_{-}$, 
one obtains the solution
to the RG equation as follows;
\beq
C_{+}(t)=c_{+}\e^{- 3\eps c_{+}c_{-}t/2}, \quad 
C_{-}(t)=c_{-}\e^{ 3\eps c_{+}c_{-}t/2}.
\eeq
Thus one finds that the global solution is given by 
\beq
\label{dwresult_1}
\bfu(t)&=&\bfW(t)=\bfW_0[A_{+}, A_{-}]+\eps\bfrho_1[A_{+}, A_{-}],
      \nonumber \\
     &=& A_{+}(t)\bfU_{+}+A_{-}(t)\bfU_{-}\nonumber \\ 
     & & -{\eps\over 8}\biggl[
       \{2A^3_{+}(t)-6A_{+}(t)A^2_{-}(t)-A^3_{-}(t)\}\bfU_{+}
       \nonumber \\ 
     & &+ \{-A^3_{+}(t)-6A^2_{+}(t)A_{-}(t)+2A^3_{-}(t)\}\bfU_{-}
       \biggl],
\eeq
where $A_{\pm}(t)=c_{\pm}{\rm exp}\{\pm \alpha t\}$
 with $\alpha = 1-3\eps c_{+}c_{-}/2$.
In this case, the unstable manifold is represented solely with 
$A_{\pm}$ and the dynamics on the manifold is given through the evolution
 of these variables. 

To compare the result  with the exact solution given in terms
of an elliptic function, we first write down the first component 
$x(t)$ of $\bfu (t)$
given in (\ref{dwresult_1});
\beq
x(t)=(1+\frac{3}{4}\eps \beta)(c_{+}\e ^{\alpha t}+c_{-}\e ^{-\alpha t})
   -\frac{\eps}{8}\{c^3_{+}\e ^{3\alpha t}+c^3_{-}\e ^{-3\alpha t}\},
\eeq
with $\beta =c_{+}c_{-}$.  Writing
 $(1+\frac{3}{4}\eps \beta)c_{\pm}$ as $c_{\pm}$, we have
\beq
\label{dwresult_2}
x(t)=c_{+}\e ^{\alpha t}+c_{-}\e ^{-\alpha t}  
  -\frac{\eps}{8}\{c^3_{+}\e ^{3\alpha t}+c^3_{-}\e ^{-3\alpha t}\},
\eeq
still with $\alpha = 1-3\eps c_{+}c_{-}/2$ up to 
$O(\eps^2)$. In fact, if one solved the 
equation in the scalar form without converting
 the  equation to the vector one and applied the operator method adopted
 in \cite{kuni95,kuni97} and explained in Appendix A of the present paper,
one would have directly reached the form given in (\ref{dwresult_2}).
We stress that  the method adopted in \cite{kuni95,kuni97}
 is more convenient in practice than that presented here.

To make the comparison with the exact solution easier, it is found
convenient to consider the case with the initial condition $x(0)=0$.
 This implies that $c_{+}=-c_{-}$.
Further putting $c_{+}(1+3\eps\beta/8)/2=C$, one has
\beq
\label{dwresult_3}
x(t)=C\sinh\alpha t-\frac{\eps}{8}C^3\sinh^3\alpha t,
\eeq
with $\alpha=1+\frac{3}{8}\eps C^2$ up to $O(\eps^2)$.

Now the exact solution with the initial condition $x(0)=0$ reads
\beq
x(t)&=&h\ {\rm cn}(\bar{\alpha}(\eps)t+K(k), k),\nonumber \\
  \ &=&-hk'F(\bar{\alpha}(\eps)t, k),
\eeq
where 
$F(\bar{\alpha}(\eps)t, k)={\rm sn}(\bar{\alpha}(\eps)t, k)/ $
${\rm dn}(\bar{\alpha}(\eps)t, k)$ with 
$\bar{\alpha}(\eps)=\sqrt{\eps/2}k/h$,
$\rm{cn}(t, k), \rm{sn}(t, k)$ and 
 ${\rm dn}(t, k)$ are  Jacobi's 
elliptic functions with  modulus $k$,
 $K(k)$ the complete 
elliptic integral of the first kind and $k'=\sqrt{1-k^2}$ the 
complementary modulus. 
The constants $k$ and $h$
 are functions of $\eps$ and  the energy $E$ of the system, 
which is assumed to be positive  here; 
\beq
h=\sqrt{\frac{1+\sqrt{1+4\eps E}}{\eps}}\simeq \sqrt{2/\eps}
   (1+\eps E/2),\quad 
k=h\sqrt{\eps/2}(1+4\eps E)^{-1/4}\simeq 1-\eps\frac{E}{2},
\eeq
and $\bar{\alpha}(\eps)\simeq 1+\eps E$.
Notice that $k$ approaches $1$ as $\eps$ goes to $0$.
If one expands $F(\bar{\alpha}(\eps)t, k)$  w.r.t. $k$ around $k=1$
 with $\bar{\alpha}(\eps)$ fixed, one has 
\beq
F(\bar{\alpha}(\eps)t, k)\simeq
  (1+\frac{\eps}{4}E)\sinh(1+\frac{3}{4}\eps E)t 
 -\frac{\eps}{4}E\sinh ^3(1+\frac{3}{4}\eps E)t,
\eeq
up to $O(\eps ^2)$.
Notice that the expansion of the elliptic functions 
at  $k=1$ is subtle; we have made the manipulation as follows;
\beq
\sinh u -\frac{\eps}{4}Eu \cosh u&=&
      \sinh u -\frac{\eps}{4}Eu\frac{d\sinh u}{du}, \nonumber \\ 
 \ \ &\simeq& \sinh(1- \frac{\eps}{4}E)u.
\eeq
Then identifying $C=-hk'(1+\frac{\eps}{4}E)=-\sqrt{2E}(1+\frac{5}{8}\eps E)$,
one reproduces the above  result (\ref{dwresult_3}). 
In Appendix B, an elementary method is presented 
to derive the approximate formula
 obtained above.  
 
\subsection{An example with a Jordan cell; Takens equation}
As examples which involve a linear operator with a Jordan cell,
we  take the Takens equation \cite{takens} in the present subsections.
 
The Takens equation is given by 
\beq
\dot{x}=y+ax^2,\quad \dot{y}=bx^2.
\eeq
Since we are interested in a slow motion in the vicinity of the origin, we 
make a scale transformation;
\beq
x=\eps^{\alpha}X,\quad y=\eps^{\beta}Y,
\eeq
where $\eps$ is a small parameter.
To make the equation to be balanced, one finds that $\alpha=\beta$; we choose
 $\alpha=\beta=1$ for simplicity. Then we end up with
\beq
\dot{X}=Y+\eps aX^2,\quad \dot{Y}=\eps bX^2.
\eeq
 
Expanding as $X=X_0+\eps X_1+\cdots,$ and 
$Y=Y_0+\eps Y_1+\cdots$, we first solve the equation around 
$t\sim t_0$ with the initial value $\bfW(t_0)=\ ^t(X(t_0), Y(t_0))$
$=\bfW_0(t_0)+\eps\bfW_1(t_0)+\cdots$, where $t_0$ is arbitrary.i
The equations in the first few orders read
\beq
\dot{X}_0&=&Y_0,\quad \dot{Y}_0=0,\\ 
\dot{X}_1&=&Y_1+aX^2_0,\quad \dot{Y}_1=bX^2_0, 
\eeq
and so on.

We take the stationary solution as the lowest order
 one to describe a slow motion on an invariant manifold, 
in accordance with the previous treatment; namely,
\beq
X_0(t;t_0)={\rm const}=C_0(t_0),\quad Y_0(t;t_0)=0,
\eeq
$C_0(t_0)$ is an integral constant.
Accordingly, $\bfW_0(t_0)=\ ^t(C_0(t_0), 0)$, i.e., the unperturbed manifold 
 M$_0$ is the $X$ axis.

Then the first order equation is solved successively from $Y_1$ to 
$X_1$ to yield
\beq
Y_1(t;t_0)&=& bC^2_0(t_0)(t-t_0)+C_1(t_0), \nonumber \\
X_1(t;t_0)&=&{b\over 2}C^2_0(t_0)(t-t_0)^2+(C_1(t_0)+aC^2_0(t_0))(t-t_0),
\eeq
where $C_1(t_0)$ is another integral constant.
Accordingly, 
$\bfW_1(t_0)=\ ^t(0, C_1(t_0))$; namely, the modification of the 
invariant manifold is given in the $Y$ direction;
\beq
{\rm M}_1=\{ (X, Y)\vert (X, Y)=(C_0, C_1)\}.
\eeq
 
Up to this order, we have
\beq
X(t;t_0)&=&C_0(t_0)+\eps\{
{b\over 2}C^2_0(t_0)(t-t_0)^2+(C_1(t_0)+aC^2_0(t_0))(t-t_0)
                 \}   ,\\ 
Y(t;t_0)&=&\eps\{  bC^2_0(t_0)(t-t_0)+C_1(t_0)\},
\eeq
with
$\bfW(t_0)=\ ^t(C_0(t_0), \eps C_1(t_0))$.

Now applying the RG equation to $X(t;t_0)$ and $Y(t;t_0)$ thus obtained, 
we have
\beq
\dot{C}_0=\eps(C_1+aC_0(t)),\quad \dot{C}_1=bC_0^2.
\eeq
The trajectory is given by 
\beq
X(t)=X(t;t)=C_0(t), \quad Y(t)=Y(t;t)=\eps C_1(t),
\eeq
which shows that the original Takens equation is reproduced.
This means that Takens equation is  "irreducible" and can not be reduced
 to a simpler equation.
 
A few comments are in order: 
In this case, the invariant manifold is represented with two variables 
 in accordance with the dimension of the Jordan cell although
 the dimension of the unperturbed solution is one; namely, the dimension of the
 invariant manifold is increased from that of the unperturbed manifold M$_0$. 
 The amplitude of the 
 trajectory in the second direction, $\eps C_1(t)$ is small 
 compared with the amplitude in the first direction $C_0$,
 while the time dependence of the second variable is large in comparison 
 with the first variable.
\setcounter{equation}{0}
\section{Generic systems with the linear operator having zero eigenvalues}

In this section, we shall examine invariant manifolds and 
slow motions given by 
 generic systems which have a linear operator having zero eigenvalues.
We shall show how uniquely the initial values are chosen
 by using a simple formula for the special solutions to differential equations
 as in the previous section; the initial values are determined 
 successively so that terms which give  fast motions
  and those proportional to the unperturbed solution do not appear. 
We call   these unwanted terms "dangerous ones".
We shall also show a necessary condition on the type of equations 
for the RG method to be applicable, which condition is relevant 
 when the linear operator has a Jordan cell.

We  treat the following rather generic  vector equations in this section:
\beq
\label{eq:ei-0}
\d _t\bfu =A\bfu+ \eps \bfF (\bfu),
\eeq
where $\d _t\bfu =\d \bfu/\d t$, 
$A$ is a linear operator, $\bfF$ a nonlinear function
of $\bfu$ and $\eps$ is a small parameter ($\vert \eps\vert<1$).
We assume that $A$ has multiply degenerated zero eigenvalues and 
other  eigenvalues of $A$ have a negative real part.

We are interested in constructing the attractive manifold M at 
$t\rightarrow \infty$ and the reduced dynamics on it.
We try to construct solve the problem
 in the perturbation theory by expanding $\bfu$ as
\beq
\bfu (t;t_0)= \bfu _0(t;t_0) +\eps \bfu_1(t;t_0) +\eps^2 \bfu_2(t;t_0) + 
\cdots,
\eeq
with the initial value $\bfW (t_0)$ at an arbitrary time $t_0$.
We suppose that the equation has been solved up to $t=t_0$ and the solution
has the value $\bfW(t_0)$ at $t_0$.  Actually, the initial value must be 
determined by the perturbative solution self-consistently;
indeed, $\bfu(t)=\bfW(t)$ is the solution to (\ref{eq:ei-0}) in 
the global domain. Therefore it
should be also expanded as follows;
\beq
\bfW(t_0)&=&\bfW_0(t_0)+\eps \bfW_1(t_0)+\eps^2\bfW_2(t_0)+\cdots,\nonumber \\ 
   &=& \bfW_0(t_0)+\bfrho(t_0),
\eeq
where $\bfrho(t_0)$ is supposed to be an independent function of $\bfW_0$.
They are not yet known at present 
 but will be determined so that the perturbative
expansion becomes valid. One of the main purposes in this section 
is how sensibly the
initial values can be determined order by order.

The equations in the first few orders read
\beq
\label{eq:ei-1}
(\d_t -A)\bfu_0&=&0, \\
\label{eq:ei-2} 
(\d_t-A)\bfu_1&=& \bfF(\bfu_0), \\
\label{eq:ei-3} 
(\d_t-A)\bfu_2&=& \bfF '(\bfu_0)\bfu_1, 
\eeq
where 
\beq
(\bfF '(\bfu_0)\bfu_1)_i=\sum_{j=1}^{n}
\left\{\d (F '(\bfu_0))_i/\d (u_0)_j\right\}(u_1)_j,
\eeq
 if $\bfu$ is an $n$-dimensional vector.

W treat the two cases separately where  $A$ has semi-simple $0$ eigenvalues 
or a Jordan cell.

\subsection{When $A$ has semi-simple zero eigenvalues}

In this subsection, we treat the case where $A$ has  semi-simple 0
 eigenvalues. Let the dimension of ker$A$ be $m$;
\beq
A\bfU_i=0, \quad (i=1, 2, \dots ,m).
\eeq
We suppose that other eigenvalues have negative real parts;
\beq
A\bfU_{\alpha}=\lambda_{\alpha}\bfU_{\alpha},
 \quad (\alpha=m+1, m+2, \cdots , n),
\eeq
where Re$\lambda_{\alpha}<0$.
One may assume without loss of generality 
that $\bfU_i$'s and $\bfU_{\alpha}$'s are linearly independent.

The adjoint operator $A^{\dag}$ has the same eigenvalues as $A$ has;
\beq
A^{\dag}\tilde{\bfU}_i&=&0, \quad (i=1, 2, \dots ,m),\nonumber\\ 
A^{\dag}\tilde{\bfU}_{\alpha}&=& \lambda^{*}_{\alpha}\tilde{\bfU}_{\alpha},
\quad (\alpha =m+1, m+2, \cdots , n).
\eeq
Here we suppose that $\tilde{\bfU}_i$' and $\tilde{\bfU}_{\alpha}$'s
 are linearly independent. Without loss of generality, one can choose
 the eigenvectors so that 
\beq
\la \tilde{\bfU}_i, \bfU_{\alpha}\ra=0=\la\tilde{\bfU}_{\alpha}, \bfU_i\ra,
\eeq
with $1\le i\le m$ and $m+1\le \alpha \le n$.
  
We denote  the projection operators by $P$ and $Q$ which projects onto
 the kernel of $A$ and the space orthogonal to ker$A$, respectively.
 The projection operators can be constructed in terms of $\bfU_i$ and
 $\tilde{\bfU}_i$ $(i=1, 2, \dots ,m)$ as follows: 
 Let $\hat{U}_P$ be an $n\times m$ matrix defined by
 $\hat{U}_P=(\bfU_1, \bfU_2, \dots , \bfU_m)$ and $\hat{\tilde{U}}_P$ by
$\hat{\tilde{U}}_P=(\tilde{\bfU}_1, \tilde{\bfU}_2, \dots ,\tilde{\bfU}_m)$,
then  
\beq
P=\hat{U}_P(\hat{\tilde{U}}^{\dag}_P\hat{U}_P)^{-1}\hat{\tilde{U}}^{\dag}_P,
\eeq
 and $Q=1-P$.

Since we are interested in the asymptotic state as $t\rightarrow \infty$,
we  may assume that the lowest-order initial value  belongs to ker$A$:
\beq
\bfW_0(t_0)=\sum_{i=1}^{m}C_i(t_0)\bfU_i=\bfW_0[\bfC].
\eeq
Thus trivially,
$\bfu_0(t;t_0)=\e^{(t-t_0)A}\bfW_0(t_0)=\sum_{i=1}^{m}C_i(t_0)\bfU_i$.
We notice that a natural parameterization of 
the invariant manifold in the lowest order M$_0$ is 
given by the set of the integral constants 
$\bfC=\ ^t(C_1, C_2, \cdots, C_m)$ being varied.

The first order equation (\ref{eq:ei-2}) 
with the initial value $\bfW_1(t_0)$ 
 which is not yet determined is formally 
solved to be
\beq
\label{eq:ei-formalsol-1}
\bfu_1(t;t_0)= 
\e^{(t-t_0)A}\bfW_1(t_0)+\int_{t_0}^{t}ds\e^{(t-s)A}\bfF(\bfu_0(s;t_0)).
\eeq
We remark that one may assume that the initial value $\bfW_1(t_0)$ is 
independent of 
$\bfW_0(t_0)$, namely $\bfW_1(t_0)$ belongs to the Q-space, because
 if $\bfW_1(t_0)$ had a component belonging to ker$A$, the component could be
 "renormalized away" into $\bfW_0$.
Inserting the identity $I=P+Q$ between the two functions in the
 integral, we have
\beq
\bfu_1(t;t_0)&=&
\e^{(t-t_0)A}[\bfW_1(t_0)+A^{-1}Q\bfF(\bfW_0(t_0))]\nonumber \\
            & & +(t-t_0)P\bfF (\bfW_0(t_0))-A^{-1}Q\bfF(\bfW_0(t_0)).
\eeq
The first term has a possibility to give rise to a fast motion, which should 
be avoided and called "dangerous" term: The "dangerous" terms here are
analogous with 
 divergent terms in quantum field theory, which are subtracted away by 
  counter terms, analogue to the initial values $\bfW_i$ here,
 self-consistently.
Indeed it is nice that the initial value $\bfW_1(t_0)$ not yet determined 
can be chosen so as to cancel out the  "dangerous" term
 as follows;
\beq
\label{eq:init-1}
\bfW_1(t_0)=-A^{-1}Q\bfF(\bfW_0(t_0)),
\eeq
which satisfies $P\bfW_1(t_0)=0$ and is a function solely of $\bfC(t_0)$.
  Thus we have for the first order solution
\beq
\label{eq:gen-sol-1}
\bfu_1(t;t_0)=(t-t_0)P\bfF -A^{-1}Q\bfF,
\eeq
where the argument of $\bfF$ is $\bfW_0[\bfC]$.
Notice that Eq.(\ref{eq:gen-sol-1}) is consistent with (\ref{eq:init-1}).
We remark that what we have done is actually a simple thing; we have 
suppressed the unperturbed part that would be  damped out as 
$t\rightarrow \infty$.

Now the invariant manifold is modified to M$_1$ given by
\beq
{\rm M}_1=\{\bfu \vert \bfu=\bfW_0-\eps A^{-1}Q\bfF(\bfW_0)\}.
\eeq

If one stops to this order, the approximate solution reads
\beq
\bfu(t;t_0)=\bfW_0+\eps\{
    (t-t_0)P\bfF -A^{-1}Q\bfF
                            \}.
\eeq
Then the RG equation $\d\bfu/\d t_0\vert_{t_0=t}=0$ gives 
\beq
\label{eq:rg-gen}
\dot{\bfW}_0(t)=\eps P\bfF(\bfW_0(t)),
\eeq
which is reduced to an $m$-dimensional coupled equation,
\beq
\label{eq:rg-gen-2}
\dot{C}_i(t)=\eps \la\tilde{\bfU}_i, \bfF(\bfW_0[\bfC])\ra,
 \quad (i=1, 2, \cdots ,m).
\eeq
The global solution representing a trajectory on the invariant manifold 
up to this order  is given by                              
\beq
\label{eq:rg-sol-1}
\bfu(t)=\bfu(t; t_0=t)=\sum_{i=1}^{m}C_i(t)\bfU_i
- \eps A^{-1}Q\bfF(\bfW_0[\bfC]),
\eeq
with $\bfC(t)$ being the solution to (\ref{eq:rg-gen-2}).

In short, we have derived the invariant manifold as 
the initial value represented by (\ref{eq:rg-sol-1})
 and the reduced dynamics (\ref{eq:rg-gen-2}) on it in the RG
  method in the first order 
 approximation.
 
The second order solution can be obtained as follows.
As  has been done in the first order case,
the second order solution is formally given by
\beq
\label{eq:gen-2}
\bfu_2(t;t_0)= 
\e^{(t-t_0)A}\bfW_2(t_0)+\int_{t_0}^{t}ds\e^{(t-s)A}\bfF '(\bfu_0(s;t_0))
\bfu_1(s;t_0).
\eeq
We may assume again that the initial value $\bfW_2(t_0)$ 
belongs to the Q-space.
A straightforward evaluation of the  integral yields
\beq
\bfu_2(t;t_0)&=&
\e^{(t-t_0)A}\left[\bfW_2(t_0)-
\left\{A^{-1}Q\bfF 'A^{-1}Q\bfF - A^{-2}Q\bfF 'P\bfF 
\right\}\right]
  \nonumber \\ 
  & & 
  +A^{-1}Q\bfF 'A^{-1}Q\bfF - A^{-2}Q\bfF 'P\bfF
  - (t - t_0)\left\{ P\bfF 'A^{-1}Q\bfF + A^{-1}Q\bfF 'P\bfF 
     \right\}\nonumber \\
  & & + \frac{1}{2}(t - t_0)^2P\bfF 'P\bfF ,
\eeq
where the argument of $\bfF$ and $\bfF'$ is $\bfW_0[\bfC]$.
The initial value can be now determined so as to cancel out the 
"dangerous term", i.e., the fast moving part, as before;
\beq
\label{eq:init-2}
\bfW_2(t_0) = A^{-1}Q\bfF '(\bfW_0)A^{-1}Q\bfF(\bfW_0)
- A^{-2}Q\bfF 'P\bfF ,
\eeq
which belongs to the Q-space.
This implies that the invariant manifold is modified to M$_2$ represented
 by 
$\bfW=\bfW_0[\bfC]+\bfrho[\bfC];\ \bfrho\simeq \eps\bfW_1+\eps^2\bfW_2$.
Thus we obtain for the second order solution
\beq
\label{eq:gen-sol-2}
\bfu_2(t;t_0) 
&=&
A^{-1}Q\bfF 'A^{-1}Q\bfF - A^{-2}Q\bfF 'P\bfF
- (t - t_0)\left\{ P\bfF 'A^{-1}Q\bfF + A^{-1}Q\bfF 'P\bfF \right
\}
\nonumber \\
&&
+ \frac{1}{2}(t - t_0)^2P\bfF 'P\bfF .
\eeq
Notice that Eq.(\ref{eq:gen-sol-2}) is consistent with (\ref{eq:init-2}).
Thus the full expression of the solution up to the second order is given by 
\beq
\bfu(t;t_0)
&=&
\bfW_0(t_0) + \eps
\{(t - t_0)P\bfF -A^{-1}Q\bfF\}
\nonumber \\ 
& &
+ \eps^2 \biggl[
A^{-1}Q\bfF 'A^{-1}Q\bfF - A^{-2}Q\bfF 'P\bfF
- (t - t_0)\left\{ P\bfF 'A^{-1}Q\bfF + A^{-1}Q\bfF 'P\bfF \right
\}
\nonumber \\
&&
+ \frac{1}{2}(t - t_0)^2P\bfF 'P\bfF
\biggl] .
\eeq                        

The RG equation $\d\bfu/\d t_0\vert_{t_0=t}=0$ reads
\beq 
\label{eq:emy1}
\dot{\bfW}_0(t) - \eps P\bfF -
 \eps A^{-1}Q\bfF '\dot{\bfW}_0
+ \eps^2\left\{
P\bfF 'A^{-1}Q\bfF + A^{-1}Q\bfF 'P\bfF \right\} = 0.
\eeq
Operating the projections $P$ and $Q$ to the both sides of \rf{eq:emy1},
respectively, we have
\begin{eqnarray} \label{eq:emy2}
\dot{\bfW}_0(t) - \eps P\bfF 
+ \eps^2P\bfF 'A^{-1}Q\bfF &=& 0, \\
\label{eq:emy3}
-\eps A^{-1}Q\bfF '\dot{\bfW}_0
+ \eps^2A^{-1}Q\bfF 'P\bfF &=& 0 .
\end{eqnarray}
Firstly, we notice that the last equation \rf{eq:emy3} is reduced to 
\[
\eps A^{-1}Q\bfF '(- \dot{\bfW}_0 + \eps P\bfF ) = 0,
\]
which is identically satisfied on account of \rf{eq:emy2} up to 
 this order.
 Thus, we end up with the reduced equation given by 
\beq
\dot{\bfW}_0(t) =\eps P\bfF - \eps ^2P\bfF 'A^{-1}Q\bfF  ,
\eeq
which is further reduced  to
\beq
\label{eq:rg-final}
\dot{C}_i=\eps\la\tilde{U}_i, \bfF-\eps\bfF'A^{-1}Q\bfF\ra, 
\eeq
 for $i=1, 2, \dots,m$.
 
The global solution giving the trajectory on the invariant manifold is given 
by the initial value as 
\beq
\bfu(t)&=&\bfW(t)=\bfW_0[\bfC]+\bfrho[\bfC]
\nonumber \\ 
&=&
\bfW_0[\bfC]-\eps A^{-1}Q\bfF
+ \eps^2 \{
A^{-1}Q\bfF 'A^{-1}Q\bfF - A^{-2}Q\bfF 'P\bfF\},
\eeq                        
with $\bfC(t)$ being the solution to (\ref{eq:rg-final}).
Notice that the argument of $\bfF$ and $\bfF'$ in the above expression 
is all $\bfW_0$, hence the r.h.s is a function of $\bfC$, i.e.,
$\bfu(t)=\bfu[\bfC]$. Recall that $\bfC$ was the integral constants of
 the unperturbed solution.

A couple of  remarks are in order:
(1) When the present formulation is applied to the 
Lorenz model\cite{lorenz}
 around the first bifurcation point which is simple one, 
the result coincides 
with that given in \cite{kuni97}; in other words, the present formulation 
gives a foundation to the prescription adopted in that paper.
(2) The present formulation using the projection operators and the
 resultant RG equation (\ref{eq:rg-final}) 
governing the slow motion resemble those by  Mori theory \cite{mori}
 for stochastic motions.
  
\subsection{When $A$ has a Jordan cell}

In this subsection, we treat the case where $A$ has a Jordan cell.
 We assume the Jordan cell is two-dimensional for simplicity, and we define 
 the normalized vectors $\bfU_1$ and $\bfU_2$ by
\beq
A\bfU_1=0, \quad A\bfU_2=\bfU_1.
\eeq
The conjugate vectors $\tilde{\bfU}_1$ and $\tilde{\bfU}_2$ satisfy
\beq
A^{\dag}\tilde{\bfU}_2=0, \quad A^{\dag}\tilde{\bfU}_1=\tilde{\bfU}_2,
\eeq
where $A^{\dag}$ is the conjugate of $A$.
The normalization condition is given by 
\beq \label{eq:ein1}
\la\tilde{\bfU}_1, \bfU _1\ra = \la\tilde{\bfU}_2, \bfU _2\ra = 1,
\la \tilde{\bfU}_1, \bfU _2 \ra = 0,
\eeq
where $\la\ ,\ \ra$ denotes the inner product.
Note that $\la \tilde{\bfU}_2, \bfU _1 \ra = 0$ automatically holds.

We denote by $P$ the projection operator to the subspace (P-space)
spanned by $\bfU_1$ and $\bfU_2$;namely, for any vector $\bfu$,
\beq
P\bfu=\alpha \bfU_1+\beta \bfU_2,
\eeq
with 
$\alpha=\la  \tilde{\bfU}_1, \bfu\ra  $ and 
$\beta=\la  \tilde{\bfU}_2,\bfu \ra$.

Let $Q$ be the projection operator to the subspace (Q-space) 
compliment of the P-space.
Then one can verify that 
\beq
\e^{tA}\bfu&=& \e^{tA}(P+Q)\bfu, \nonumber \\ 
  \        &=& (\alpha +\beta t)\bfU_1+\beta\bfU_2+\e^{tA}Q\bfu.
\eeq
So much for the preliminaries.

Now let us proceed to obtain the asymptotic solution to (\ref{eq:ei-1})
 as $t\rightarrow \infty$ by the perturbation method: 
Since we are interested in constructing the invariant manifold, 
let us take the stationary solution as the lowest order one;
\beq
\bfu_0(t;t_0)=C_0(t_0)\bfU_1 ,
\eeq
accordingly, the initial value reads
\beq
\bfW_0(t_0)=C_0(t_0)\bfU_1.
\eeq
Notice that we have not included a component in $\bfU_2$ direction.
The lowest order manifold is 
\beq
{\rm M}_0=\{\bfu\vert \bfu=C_0\bfU_1\}.
\eeq
                 
The first order solution is formally given by (\ref{eq:ei-formalsol-1}).
The first order initial value is chosen to be independent of $\bfW_0$;
\[
\bfW _1(t_0) = C_1(t_0)\bfU _2 + Q\bfW _1(t_0) .
\]
A simple evaluation of the integral in  (\ref{eq:ei-formalsol-1})
gives the first order solution as 
\beq
\label{eq:jor-first}
\bfu_1(t;t_0) &=&
\e^{(t-t_0)A}[Q\bfW_1(t_0)+A^{-1}Q\bfF]\nonumber \\ 
   & & +  
\{C_1(t_0)(t - t_0) + \alpha_F(t-t_0)+\beta _F{1 \over 2}(t-t_0)^2\} \bfU_1
\nonumber \\ 
& & +\{ C_1(t_0) + \beta_F(t-t_0) \} \bfU_2 -A^{-1} Q\bfF+ 
O((t - t_0)^{n\geq 2}),
\eeq
where 
\beq
\alpha_F=\la  \tilde{\bfU}_1, \bfF\ra  ,  \quad 
\beta_F=\la  \tilde{\bfU}_2, \bfF\ra  .
\eeq
The argument of $\bfF$ in the above expressions is $\bfW_0[C_0]$.
The initial value can be determined so as to cancel out the fast mode
 as before, namely,
\beq
Q\bfW_1(t_0)=-A^{-1}Q\bfF,
\eeq
which implies that the invariant manifold is modified to
\beq
{\rm M}_1=\{\bfu \vert \bfu=C_0\bfU_1+\eps C_1\bfU_2-
\eps A^{-1}Q\bfF \}.
\eeq
Then the solution up to the first order is obtained as 
\beq
\bfu(t;t_0)
&=&
C_0(t_0)\bfU _1 + \eps\biggl[\{\alpha_F(t - t_0) 
+ C_1(t_0)(t - t_0) + \frac{1}{2}\beta_F(t - t_0)^2\}\bfU_1
\nonumber \\ 
& &  
+ \{ \beta_F(t-t_0) + C_1(t_0) \} \bfU_2 - A^{-1}Q\bfF\biggl].
\eeq

The RG equation in this order is given by
\beq
0 = \dot{C}_0\bfU_1 - 
\eps \{ (\alpha_F + C_1)\bfU_1 + (\beta_F - \dot{C}_1)\bfU_2
+ \frac{1}{A}Q\bfF '\dot{C}_0\bfU_1 ,
\eeq
which leads to 
\beq
\label{eq:rg-gen-jor}
\dot{C}_0 = \eps \left(\la  \tilde{\bfU}_1, \bfF\ra   + C_1\right),
  \quad 
\dot{C}_1= \la  \tilde{\bfU}_2, \bfF\ra  .
\eeq
We now see that the trajectory in the invariant manifold M$_1$ is 
 given by 
\beq
\bfu(t)&=&\bfW(t)\simeq \bfW_0[C_0]+\bfW_1[C_0],\nonumber \\ 
   &=&C_0(t)\bfU_1+ \eps C_1(t)\bfU_2-\eps A^{-1}Q\bfF,
\eeq
with $C_0(t)$ and $C_1(t)$ being governed by (\ref{eq:rg-gen-jor}).
Notice that $\bfu(t)$ is a functional of $C_0(t)$ and $C_1(t)$.

The second order solution is obtained, similarly. The formal solution to the
second order equation is given by Eq.(\ref{eq:gen-2}). 
Let $\bfg = C_1(t_0)\bfU _1 + \alpha _F\bfU _1 + \beta _F\bfU _2$
and $\bfh = -\dfrac{1}{A}Q\bfF + C_1\bfU _2$.
Then, a simple manipulation as before gives 
\beq
\bfu_2(t;t_0) 
&=&
\e^{(t-t_0)A}[\bfW_2(t_0)+\{A^{-1}Q\bfF '\bfh +A^{-2}Q\bfF'\bfg \}]
   \nonumber \\
 & & + (t - t_0)\left\{
-A^{-1}Q\bfF'\bfg + P\bfF '\bfh \right\}
-\{A^{-1}Q\bfF '\bfh +A^{-2}Q\bfF'\bfg \}\nonumber \\ 
  & & 
+ O((t - t_0)^2) .
\eeq
Thus  the initial value is determined to be
\beq
\bfW_2(t_0) = -\left(A^{-1}Q\bfF '\bfh +A^{-2}Q\bfF'\bfg\right),
\eeq
so that the fast modes disappear; notice that $\bfW_2$ belongs to 
the Q-space.
The invariant manifold is modified now in an apparent way;
 so we do not write the specification of the second order manifold 
 M$_2$.
Hence we have 
\beq
\label{eq:jordan-sol-2}
\bfu_2(t;t_0) &=&
(t - t_0)\left\{
-A^{-1}Q\bfF'\bfg + P\bfF '\bfh \right\}
-\{A^{-1}Q\bfF '\bfh +A^{-2}Q\bfF'\bfg \}\nonumber \\ 
  & &  + O((t - t_0)^2) .
\eeq
Applying the RG equation to $\bfu=\bfu_0+\eps\bfu_1+\eps^2\bfu_2$ thus
obtained, we have
\beq
\dot{C}_0\bfU _1 - \eps (\alpha _F + C_1)\bfU _1
+ \eps (\dot{C}_1 - \beta _F)\bfU _2
- \eps A^{-1}Q\bfF ' \dot{C}_0\bfU _1
\nonumber && \\ 
- \eps ^2\left\{ -A^{-1}Q\bfF'\bfg + P\bfF '\bfh \right\}
- \eps ^2\left\{
A^{-1}Q\bfF '\dot{C}_1\bfU _2 + A^{-2}Q\bfF '\dot{C}_1\bfU _1
\right\} &=& 0
\label{eq:emy4}
\eeq
Operating $P$ and $Q$ to \rf{eq:emy4}, we obtain 
\begin{eqnarray} 
0&=&\dot{C}_0\bfU _1 + \eps \dot{C}_1\bfU _2
- \eps \bfg - \eps ^2P\bfF '\bfh
, \label{eq:emy5} \\
0&=&
-\eps ^2A^{-2}Q\bfF '\dot{C}_1\bfU _1
 - \eps A^{-1}Q\bfF '\left\{
 \dot{C}_0\bfU _1 + \eps \dot{C}_1\bfU _2 - \eps \bfg
                     \right\}.
\label{eq:emy6}
\end{eqnarray}
We remark that from \rf{eq:emy5} and \rf{eq:emy6},
\beq
\label{jor;comp}
Q\bfF'\bfU _1 = 0
\eeq
must hold as a compatibility condition, which gives
 a necessary condition for the RG method to work
for higher approximations.
With this condition taken for granted, 
$\bfg$ is reduced to
$\bfg=\beta_F\bfU_2$.

Equating the components in the $\bfU_1, \bfU_2$ in \rf{eq:emy5}, we have
the reduced dynamics as follows,
\beq
\dot{C}_0 
&=& 
\eps (\la  \tilde{\bfU}_1, \bfF+\eps\bfF'\bfh \ra   + C_1), \\
\dot{C}_1
&=&
 \la  \tilde{\bfU}_2, \bfF + \eps \bfF'\bfh \ra   ,
\eeq
where the argument of $\bfF$ and $\bfF '$ is 
$\bfW_0(t)=C_0(t)\bfU_1$.

The trajectory on the manifold M$_2$ is given by 
\beq
\bfu(t)&=&\bfW(t)=\bfW_0(t)+\eps\bfW_1(t)+\eps^2\bfW_2(t),\nonumber \\
       &=&C_0(t)\bfU_1+ \eps C_1(t)\bfU_2 - \eps A^{-1}Q\bfF
       \nonumber \\ 
       & & -\eps^2\{ A^{-1}Q\bfF '\bfh+ 
        \la\tilde{\bfU}_2, \bfF\ra A^{-2}Q\bfF'\bfU_2 \}.
\eeq
Comments are in order: The solution is solely described  by the 
coordinates $C_0$ and $C_1$ representing the P space as in the previous 
subsection. Conversely, the dynamics can not be described 
 only by the coordinate in the zero-th manifold in this case; the dimension
 of the invariant manifold is increased from that of the unperturbed
  invariant manifold.
We remark that the  formulae obtained above are completely consistent
 with those given for the Takens equation in \S 3.

\subsection{Practical way}
We have formulated the RG method so that it is clarified that the method
 concerns with 
 the initial values and the invariant manifold is constructed as 
 the initial value perturbatively: The initial values are determined
 so that terms proportional to the unperturbed solution and those 
 representing fast motions disappear in the perturbed solutions.
 In effect, the special solutions of the higher order equations are 
 composed of secular terms which are proportional to the unperturbed
 solution and vanish at $t=t_0$ and  solutions independent of the 
 unperturbed solution.  We remark that this way of construction
  of the perturbative special solutions have been adopted 
  in \cite{kuni95,kuni97,qm}.
 In this subsection, having known the above fact, we  present
 the rules  for constructing
  the special solutions in the form of an operator method;
 a detailed account of this method
 is given  in Appendix A. 
  This subsection will  constitute a practical
  summary of the results obtained in the previous subsections.
 
When we try to obtain an invariant manifold and the reduced dynamics on it,
the solution to the unperturbed equation (\ref{eq:ei-1}) is given by a
stationary one
\beq
\bfu_0(t;t_0)=\bfW_0(t_0).
\eeq
Then the first order solution is given by 
\beq
\bfu_1(t;t_0)&=&\frac{1}{\d_t-A}\bfF (\bfu_0)
              =\frac{1}{\d_t-A}(P\bfF (\bfu_0)
                +Q\bfF(\bfu_0)),\nonumber \\ 
              &=&(t-t_0)P\bfF+ {1\over {-A}}Q\bfF.
\eeq
Here (\ref{app;1}) and (\ref{app;2}) have been used and $A^{-1}$ is written 
as $\frac{1}{A}$.

Similarly, the second order solution is given by
\beq
\bfu_2(t;t_0)&=&\frac{1}{\d_t-A}\bfF '(\bfu_0)\bfu_1,\nonumber \\ 
             &=&\frac{1}{\d_t-A}(P+Q)\bfF' (\bfu_0)\{
                (t-t_0)P\bfF+ {1\over {-A}}Q\bfF\},\nonumber \\ 
             &=&{1\over 2}(t-t_0)^2P\bfF'P\bfF
                +(t-t_0)P\bfF'{1\over {-A}}Q\bfF
               \nonumber \\ 
              & &\ \ 
              -\{(t-t_0){1\over{A}}+ {1\over {A^2}}\}Q\bfF'P\bfF
               +{1\over{-A}}Q\bfF'{1\over {-A}}Q\bfF,
\eeq
which coincides with (\ref{eq:gen-sol-2}). Here we have used the formulae
 (\ref{app;3}), (\ref{app;4}) and (\ref{app;5}).
  The efficiency of the operator method is apparent.
 
Next, let us consider the case where $A$ has a two-dimensional Jordan cell.
We take a stationary  solution as the zeroth-order one;
\beq
\bfu_0(t;t_0)=C_0(t_0)\bfU_1.
\eeq

Notice that the kernel of $A$ is not yet fully spanned by the solution.
Then the first order solution is given by a sum of the remaining component of 
ker$A$ and the special solution;
\beq
\bfu_1(t;t_0)&=&\frac{1}{\d_t-A}\{0\}+
                \frac{1}{\d_t-A}\bfF (\bfu_0),\nonumber \\ 
             &=&C_1(t_0)(t-t_0)\bfU_1+C_1(t_0)\bfU_2+
             \frac{1}{\d_t-A}(P\bfF (\bfu_0)+Q\bfF(\bfu_0)),\nonumber \\ 
              &=&C_1(t_0)(t-t_0)\bfU_1+C_1(t_0)\bfU_2\nonumber \\
              &\ \ &
              +\{(t-t_0)P\bfF + 
              {1\over 2}(t-t_0)^2\la  \tilde{\bfU}_2,\bfF\ra  \bfU_1\}+ 
              {1\over {-A}}
              Q\bfF,
\eeq
which is found to coincide with Eq.(\ref{eq:jor-first}).
Here (\ref{app;2}) and (\ref{app;6})
 have been used. The unperturbed solution
 $\frac{1}{\d_t-A}\{0\}$ may be obtained in the following way:
Putting $\frac{1}{\d_t-A}\{0\}=a(t)\bfU_1+b(t)\bfU_2$, one has a coupled 
equation $\dot{a}=b,\ \ \dot{b}=0$, which is solved to yield
$a(t)=C_1(t_0)(t-t_0),\ \ b(t)=C_1(t_0)$.

The second order solution is given by
\beq
\bfu_2(t;t_0)&=& \frac{1}{\d_t-A}\bfF'(\bfu_0)\bfu_1(t;t_0),\nonumber \\ 
             &=& \frac{1}{\d_t-A}(P+Q)\bfF'(\bfu_0)\bfu_1(t;t_0).
\eeq
Then the r.h.s. is a sum of terms which are in the form
$\frac{1}{\d_t-A}(t-t_0)^nP\bfG(\bfu_0)$ or 
$\frac{1}{\d_t-A}(t-t_0)^nQ\bfG(\bfu_0)$, which are calculated in Appendix A.
Actually, since 
terms proportional to $(t-t_0)^n$ ($n\ge 2$) do not contribute to the 
RG equation nor to the resulting trajectory,
one needs not calculate the terms of the form 
$\frac{1}{\d_t-A}(t-t_0)^nP\bfG(\bfu_0)$ with $n\ge 1$.
Thus we easily reach the final result given in (\ref{eq:jordan-sol-2}).

\subsection{Normal form}

As a simple example, we try to reduce the following
two-dimensional  evolution equation with a Jordan cell;
\beq
\dot{x}=y +a_1x^2+b_1xy+c_1y^2, 
\quad \dot{y}=a_2x^2+b_2xy+c_2y^2,
\eeq
with $a_i$, $b_i$ and $c_i$ ($i=1, 2$) being constant. 
As in the Takens equation examined in \S 3, we make a scale transformation
\beq
x=\eps X,\quad y=\eps Y.
\eeq
Then the equation is reduced to 
\beq
(\d_t-A)\bfu=\eps\bfF(\bfu),\quad \bfu= \ ^t(X, Y),
\eeq
where
\beq
A=\pmatrix{0\ 1\cr 
   0\ 0},\quad
   \bfF(\bfu)= {a_1X^2+b_1XY+c_1Y^2 \choose a_2X^2+b_2XY+c_2Y^2}.
\eeq   
Here the P-space is spanned by 
\beq
\bfU_1={1\choose 0},\quad {\rm and}\quad \bfU_2={0\choose 1},
\eeq
which satisfy  $A\bfU_1=0$ and $A\bfU_2=\bfU_1$. And 
$\tilde{\bfU}_i=\bfU_i$ $(i=1, 2)$.

Expanding as $\bfu=\bfu_0+\eps \bfu_1+\cdots,$ with
 $\bfu_i=\ ^t(X_i, Y_i)$,
 we first solve the equation around 
$t\sim t_0$ with the initial value 
$\bfW(t_0)=\bfW_0(t_0)+\eps\bfW_1(t_0)+\cdots$.

We take the stationary solution as the lowest order
\beq
\bfu_0(t;t_0)=A_0 (t_0)\bfU_1,
\eeq
where $A_0 (t_0)$ is an integral constant.
Accordingly, $\bfW_0(t_0)=\ ^t(A_0 (t_0), 0)$; the unperturbed manifold
is the $X$ axis.

According to the general argument given in \S 4.2,
we need to calculate the following quantities;
\beq
\alpha_F=\la  \tilde{\bfU}_1, \bfF(\bfu_0)\ra  =a_1A^2_0(t_0),
\quad \beta_F=\la  \tilde{\bfU}_2,\bfF(\bfu_0)\ra  =a_2A^2_0(t_0).
\eeq
Notice that $Q\bfF(\bfu_0)=0$, identically.
Then one has
\beq
\bfu_1(t;t_0)&=& 
    \{A_1(t_0)(1+a_1A_1(t_0))(t-t_0)+
{a_2\over 2}A^2_0(t_0)(t-t_0)^2\}\bfU_1\nonumber \\
    & & +\{A_1(t_0)+a_2A^2_0(t_0)(t-t_0)\}\bfU_2,
\eeq
where $A_1(t_0)$ is another integral constant.
Accordingly,
$\bfW_1(t_0)=\ ^t(0, A_1(t_0))$.
Then 
\beq
{\rm M}_1=\{\bfu \vert \bfu=\ ^t(A_0 , \eps A_1)\}.
\eeq
If we stop at this order, the RG equation reads
\beq
\dot{A}_0=\eps(A_1+a_1A^2_0), \quad \dot{A}_1=a_2A^2_0.
\eeq
This is the Takens equation; notice the trajectory is given by
$x(t)=\eps X(t;t)=\eps A_0 (t), \, y(t)=\eps Y(t;t)= \eps^2 A_1(t)$.
It means that the RG method gives 
mechanically the normal form \cite{wiggins} 
of the reduced equation. This is confirmed by proceeding to the
 second order.

In the present case, $\bfg=0$ and $\bfh=A_1(t_0)\bfU_2$, and
\beq
P\bfF'(\bfu_0)\bfh=A_1\{b_1A_0 \bfU_1+b_2A_0 \bfU_2\}.
\eeq
Then the second order solution is found to be 
\beq
\bfu_2(t;t_0)=(t-t_0)A_1\{b_1A_0 \bfU_1+b_2A_0 \bfU_2\}+ O((t-t_0)^{n\ge 2}),
\eeq
Accordingly, $\bfW_2(t_0)=0$.

Up to this order, we have $\bfu=\bfu_0+\eps\bfu_1+\eps^2\bfu_2$.
Then the RG equation reads
\beq
\dot{A}_0&=&\eps A_1+\eps(a_1A^2_0+\eps b_1A_0 A_1),\nonumber \\ 
\dot{A}_1&=&a_2A^2_0+\eps b_2A_0 A_1,
\eeq
which is the normal form when the linear matrix is of Jordan 
form\cite{wiggins}.
The trajectory is given by 
$x(t)=\eps A_0 (t), \, y(t)=\eps^2 A_1(t)$.
Thus one sees that the RG equation gives the normal form of the
 reduced evolution equation on the invariant manifold.
We remark that when $a_1=b_1=0$, the RG equation is nothing but the
 Bogdanov equation\cite{bogdanov}.

We remark that one needs the second order solution to give  the 
Bogdanov  equation, while the Takens equation is obtained in the first
approximation.
                  
\subsection{An extended Takens equation}

As the final example, 
we deal with an extension of the Takens equation to a system 
with three-degrees of freedom;
\beq
\label{et;1}
\dot{x}=y+\eps ax^2,\quad
\dot{y}=\eps bx^2,\quad 
\dot{z}=-z +\eps f(x,y,z),
\eeq 
where $f(x, y,z)$ is analytic function of $(x,y,z)$. 
We shall show the
compatibility condition (\ref{jor;comp}), 
which becomes relevant only when 
 the system has more than two-degrees of freedom,
  gives a restriction to the form of $f(x,y,z)$.

With  $\bfu=\ ^t(x, y, z)$, (\ref{et;1}) is converted to 
\beq
\label{et;2}
(\d_t-A)\bfu=\eps \bfF(\bfu),
\eeq
where
\beq
\label{et;3}
A=\pmatrix{0\ 1\ 0\cr
           0\ 0\ 0\cr
           0\ 0\ -1\cr
           },
           \quad
\bfF(\bfu)=\pmatrix{ax^2\cr
                    bx^2\cr
                    f(x,y,z)
                    }.
\eeq
Notice that $A$ has a two-dimensional Jordan cell;
\beq
A\bfU_1=0,\quad A\bfU_2=\bfU_1, \quad A\bfU_3=-\bfU_3,
\eeq
where $\bfU_1=\ ^t(1,0,0), \bfU_2=\ ^t(0,1,0), \bfU_3=\ ^t(0,0,1)$.
The projection operator to the subspace $\{\bfU_1, \bfU_2\}$ is 
given by $P={\rm diag}(1, 1, 0)$, while $Q=1-P={\rm diag}(0,0,1)$.
We also notice that $\tilde{\bfU}_i=\bfU_i$ ($i=1, 2$) in this simple 
example.
We are interested in the asymptotic behavior of the solution at 
$t\rightarrow \infty$. 
We first solve (\ref{et;2}) around $t\sim t_0$ with 
the initial value $\bfW(t_0)$ at $t=t_0$ by the perturbation theory.
The solution may be written as $\bfu(t;t_0,\bfW(t_0))$, which is expanded
as $\bfu=\bfu_0+\eps\bfu_1+\eps^2\bfu_2+\cdots.$  The initial value which is 
determined self-consistently with $\bfu$ is also expanded as 
$\bfW=\bfW_0+\eps\bfW_1+\eps^2\bfW_2+\cdots.$

When $t\rightarrow \infty$, we may take 
the stationary solution as an asymptotic one to the lowest order equation;
\beq
\label{et;4}
\bfu_0(t;t_0)=C_0(t_0)\bfU_1.
\eeq
Accordingly the initial value reads $\bfW_0(t_0)=C_0(t_0)\bfU_1$, 
which implies that the unperturbed invariant manifold is given by
\beq
\label{et;5}
{\rm M}_0=\{\bfu\vert \bfu=C_0\bfU_1\},
\eeq
namely the $x$ axis.

According to the general formulation given in the previous sections,
to obtain the first order solution, we only have to evaluate 
\beq
\label{et;6}
\alpha_F&=&\la\tilde{\bfU}_1, \bfF(\bfu_0)\ra=aC^2_0(t_0), \quad
\beta_F=\la\tilde{\bfU}_2, \bfF(\bfu_0)\ra=bC^2_0(t_0),\nonumber \\ 
 A^{-1}Q\bfF(\bfu_0)&=&-f(\bfu_0)\bfU_3.
\eeq
Then we end up with 
\beq
\label{et;7}
\bfu_1(t;t_0)&=&\{aC^2_0(t_0)(t-t_0)+C_1(t_0)(t-t_0)+{1\over 2}bC^2_0(t_0)
               (t-t_0)^2\}\bfU_1\nonumber \\ 
             & & +\{bC^2_0(t_0)(t-t_0)+C_1(t_0)\}\bfU_2+f(\bfu_0)\bfU_3.
\eeq
Accordingly, 
\beq
\bfW_1(t_0)=C_1(t_0)\bfU_2+f(\bfu_0)\bfU_3,
\eeq
which implies that the modified invariant manifold is given by
\beq
{\rm M}_1=\{\bfu \vert \bfu=\ ^t(C_0, C_1, f(C_0,0,0))\}.
\eeq
  
If we stop at this order,  the full solution is given by 
$\bfu\simeq \bfu_0+\eps\bfu_1$.  Applying the RG equation, we have
\beq
\label{et;rg}
\dot{C_0}=\eps(aC^2_0+C_1) ,\quad \dot{C_1}=bC^2_0.
\eeq
And the trajectory on the manifold  M$\simeq$M$_1$ is given by
\beq
\bfu(t)=C_0(t)\bfU_1+\eps C_1(t)\bfU_2+\eps f(C_0(t), 0, 0)\bfU_3.
\eeq

To go to the second order, we first need  to examine the 
compatibility condition;
\beq
0=Q\bfF'(\bfu_0)\bfU_1
    =\frac{\d f(x,y,z)}{\d x}\biggl\vert _{\bfu=\bfu_0}\bfU_3,
\eeq
with $\bfu_0=\ ^t(C_0, 0, 0)$, which means that when $y=z=0$,
$f(x,y,z)$ does not depend on $x$.

With this condition assumed, we can proceed to the second order. 
To obtain the second order RG equation, 
we notice the following;
$\bfh=-A^{-1}QF(\bfu_0)+C_1\bfU_2$
$=f(\bfu_0)\bfU_3+C_1\bfU_2$ and hence
$\bfF'(\bfu_0)\bfh= 0$. 
Thus the RG equation which 
gives the evolution equation of $C_{0,1}(t)$ is not modified by the 
second order perturbation. 

To obtain the second order correction to the trajectory, we need to evaluate
 the following;
\beq
-A^{-1}Q\bfF'\bfh&=&
         (f(\bfu_0)\frac{\d f}{\d z}+C_1\frac{\d f}{\d y})\bfU_3,
         \nonumber \\
-A^{-2}Q\bfF'\bfU_2&=&
                 -\frac{\d f}{\d y}\bfU_3 ,
\eeq
where the derivatives are evaluated at $\bfu=\bfu_0=\ ^t(C_0(t),0,0)$.
Thus the second order correction of the initial value reads
\beq
\bfW_2(t_0)=(f(\bfu_0)\frac{\d f}{\d z}+C_1\frac{\d f}{\d y}
-\beta_F\frac{\d f}{\d y})\bfU_3.
\eeq
Hence the trajectory in the second order approximation
is given by 
\beq
\bfu(t)&=&\bfW(t)=C_0(t)\bfU_1+\eps C_1(t)\bfU_2+\eps f(\bfu_0)\bfU_3
        \nonumber \\ 
        & & +\eps^2
        (f(\bfu_0)\frac{\d f}{\d z}+C_1\frac{\d f}{\d y}
-\beta_F\frac{\d f}{\d y})\bfU_3.
\eeq
 Here $C_{0,1}(t)$ are governed by the
Takens equation (\ref{et;rg}).  We see that 
 the higher order terms does not affect the dynamics but modifies
 the trajectory only in the $\bfU_2$ and the Q-direction.
 
Similar discussions can be made for the following  extended
Bogdanov equation,
\beq
\dot{x}=y,\quad \dot{y}=ax^2+bxy,\quad \dot{z}=-z+\eps f(x,y,z),
\eeq
where $f(x,y,z)$ is an analytic function.

\subsection{Discussion}

 We have applied the naive perturbative expansion
as the starting point of the RG method. However, the naive perturbation
expansion is not always a good starting point. There are cases where a 
scaling transformation is needed to convert the unperturbed equation to
a non-Jordan form before applying the perturbative expansion.
Let us take the following example;
\beq
\dot{\bfu}=\pmatrix{0  \ 1-\eps\cr
                 \eps \ \,\,\, 0}\bfu,
\eeq
with $ \bfu =\ ^t(x, y)$. The exact solution reads
\beq
\bfu(t)=A\e^{\lambda t}\bfU_{+}+B\e^{-\lambda t}\bfU_{-},
\eeq
where $\lambda=\sqrt{(1-\eps)\eps}$ and 
$\bfU_{\pm}=\ ^t(1,\pm\sqrt{\eps/(1-\eps)})$; i.e.,
\beq
x(t)=A\e^{\lambda t}+B\e^{-\lambda t}, \ \ 
y(t)=\sqrt{\eps/(1-\eps)}\cdot(
A\e^{\lambda t}-B\e^{-\lambda t}).
\eeq

Due to the appearance of the singular term $\sqrt{\eps}$, the naive 
perturbation does not give a sensible result even in the RG method.
In this case, we first try to convert the equation so that the converted
equation has no Jordan nature in the unperturbed part. This can be 
performed by a scale transformation;
$x=\eps^{\alpha}\xi, \ \ y=\eps^{\beta}\eta, \ \  t=\eps^{\nu}\tau$.
To make the unperturbed part to be a non-Jordan form, we choose that
$\beta-\alpha+\nu=1+\alpha-\beta+\nu=0$, which is satisfied with
$\alpha=0, \ \  \beta=-\nu=1/2$, i.e.,
$x=\xi, \ \ y=\sqrt{\eps}\eta, \ \ t=\tau/\sqrt{\eps}$.
The converted equation reads
\beq
\frac{d\xi}{d\tau}=\eta -\eps \eta, \ \ \ \frac{d\eta}{d\tau}=\xi,
\eeq
which can be now solved by the RG method with no difficulty. 
The result obtained by the RG method up to $O(\eps^2)$ reads
\beq
x(t)=A\e^{\lambda't}+B\e^{-\lambda't},\ \ 
y(t)=\sqrt{\eps}(1+\eps/2)(A\e^{\lambda't}-B\e^{-\lambda't}), 
\eeq
with $\lambda'=\sqrt{\eps}(1-\eps/2)$.
\newpage
\setcounter{equation}{0}
\section{Applications I}
In this section, we present examples of non-linear equations
 for which the unperturbed linear
 operator has no zero eigenvalues but a pair of eigenvalues $\lambda_i$
 ($i=1, 2$);
(i) $\lambda_{1,2}=\pm i\omega$ and (ii) $\lambda_{1,2}=\pm \lambda$
 where $\omega$ and $\lambda$ are real numbers.
The first case is discussed in  \cite{kuni95,kuni97} as well as 
 in \S 3. We present it here for completeness.

\subsection{Brusselator}

This is an example of systems showing  a Hopf bifurcation. 
An RG treatment of 
generic systems with a bifurcation has been given in \cite{kuni97} where
an emphasis is put on the relation of the RG method with the envelope
theory. Here we treat this interesting example in the 
present formulation emphasizing the  aspect of the RG method as the one 
to construct attractive manifolds.

The Brusselator is given by 
\beq
\frac{\d X}{\d t}&=&A-(B+1)X+X^2Y+D_X \frac{\d^2X}{\d x^2},\nonumber \\  
\frac{\d Y}{\d t}&=&BX-X^2Y+D_Y\frac{\d^2Y}{\d x^2}, 
\eeq
where $A(>0), B(>0), D_X$ and $D_y$ are constant.
We here treat a uniform system, hence the terms with the spatial 
derivatives vanish. The steady state is given by
$(X_0, Y_0)=(A, B/A)$.
Shifting the variables as
$\xi=X-X_0, \ \ \eta =Y-Y_0$,
and defining $\bfu =\ ^t(\xi, \eta)$, we have
\beq
\frac{d }{d t}\bfu =\pmatrix{(B-1)\xi+A^2\eta\cr
                     -B\xi-A^2\eta}
                     +f(\xi,\eta)\pmatrix{1\, \cr
                                          -1},
\eeq
with
\beq
f(\xi, \eta)=B/A\cdot \xi^2+2A\xi\eta +\xi^2\eta.
\eeq
The linear stability analysis shows that when $B$ exceeds the critical value
$B_c=1+A^2$,
there arises a bifurcation.

Now let us analyze the slow motion and the slow manifold around the
bifurcation (critical) point.
We define the following variables
\beq
\mu=(B-B_c)/B_c, \ \eps=\sqrt{\vert \mu\vert},  {\rm and}\ \ 
\chi={\rm sgn}(\mu),
\eeq
 accordingly, $\mu=\chi\eps^2$.
We first expand $\bfu$ and the initial value 
$\bfu(t=t_0; t_0)=\bfW(t_0)$ as Taylor series w.r.t $\eps$:
$\bfu=\eps\bfu_1+\eps^2\bfu_2+\eps^3\bfu_3+\cdots$, and
$\bfW=\eps\bfW_1+\eps^2\bfW_2+\eps^3\bfW_3+\cdots$.

The first order equation reads
\beq
(\d_t-L_0)\bfu_1={\bf 0},
\eeq
with
\beq
L_0=\pmatrix{A^2\quad \, \, \, \, \, \, \ \,\, A^2 \cr
            -(A^2+1)\ -A^2
            }.
\eeq
The solution is readily found to be
\beq
\bfu(t;t_0)=C(t_0)\bfU\e^{i\omega t}+ {\rm c.c.},
\eeq
with $\omega=A$ and
\beq
\bfU=\pmatrix{1\,\,\cr
                i\frac{1+iA}{A}
                }.
\eeq
Here $C$ is the (complex) integral constant.
Accordingly,
\beq
\bfW_1(t_0)=C(t_0)\bfU\e^{i\omega t_0}+ {\rm c.c.}.
\eeq

A simple manipulation using the formulae given in Appendix A 
gives the higher order terms as follows;
\beq
\bfu_2(t;t_0)=
\{C^2\bfV_{+}\e^{2i\omega t}+c.c.\}+\vert C\vert ^2\bfV_0,
\eeq
where
\beq
\bfV_{+}=\frac{1+iA}{3A^3}\pmatrix{-2iA\, \cr
                                   1+2iA},
\ \ \ 
\bfV_{0}=2\frac{A^2-1}{A^3}\pmatrix{0 \cr
                                    1},
\eeq
and
\beq
\bfu_3(t;t_0)=\biggl[\frac{{\cal C}_1}{2}\{(t-t_0)\bfU+
               \frac{1}{2i\omega}\bfU^{*}\}\e^{i\omega t}
                +\frac{{\cal C}_3}{4i\omega}(\bfU+\frac{1}{2}\bfU^{*})
                 \e^{3i\omega t}\biggl]+ c.c. ,
\eeq
where
\beq
{\cal C}_1&=&\chi B_c C+\{-\frac{2+A^2}{A^2}+i\frac{-4A^4+7A^2-4}{3A^3}\}
                       \vert C\vert ^2C, \\ 
{\cal C}_3&=&\{2B_cV_{+\xi}/A+2A(V_{+\eta}+V_{+\xi}\bar{U}_{\eta}+U_{\eta})
               \}C^3,
\eeq
with $V_{+\xi}$ being the $\xi$-component of $\bfV_{+}$ and so on.
Here the initial values have been chosen to be 
\beq
\bfW_2(t_0)&=&\{C^2\bfV_{+}\e^{2i\omega t_0}+c.c.\}+\vert C\vert ^2\bfV_0,
   \nonumber \\ 
\bfW_3(t_0)&=&   \biggl[\frac{{\cal C}_1}{4i\omega}\bfU^{*}\e^{i\omega t}
                +\frac{{\cal C}_3}{4i\omega}(\bfU+\frac{1}{2}\bfU^{*})
                 \e^{3i\omega t}\biggl]+ {\rm c.c.}.
\eeq                 
Here $\bfU^{*}$ is the complex conjugate of $\bfU$.

Collecting all the terms thus obtained to have the approximate $\bfu(t;t_0)$
 and applying the RG equation to it, we have
$dC/dt-\eps^2{\cal C}_1=0$, or
\beq
\frac{dC}{dt}=\alpha C+\beta \vert C\vert^2C,
\eeq
with 
\beq
\alpha=\chi (1+A^2), \ \ \beta=-\frac{2+A^2}{A^2}+
       i\frac{-4A^4+7A^2-4}{3A^3}.
\eeq
The attractive manifold is given by the initial value as 
\beq 
\bfu(t)&=&\bfW (t)\simeq 
\eps \bfW_1(t)+\eps^2\bfW_2(t)+\eps^3\bfW(t), \nonumber \\ 
 \ \    &=& \eps \{C(t)\bfU\e^{i\omega t}+{\rm c.c.}\}+
        \eps^2[(C^2(t)\bfV_{+}\e^{2i\omega t}
           +c.c.)+\vert C\vert ^2\bfV_0] \nonumber \\ 
 \ \    & & + \eps^3\biggl[[
           \frac{{\cal C}_1(t)}{4i\omega}\bfU^{*}\e^{i\omega t}
           +\frac{{\cal C}_3(t)}{4i\omega}(\bfU +\frac{1}{2}\bfU^{*})
                \e^{3i\omega t}] +{\rm c.c.}\biggl].
\eeq        
These results coincide with those obtained in the reductive perturbation
 method\cite{dissipative}.

\subsection{Unstable motion in  Lotka--Volterra system}
 The Lotka-Volterra (LV) equation\cite{lv}
  is known to be integrable, although the 
  exact analytic solutions of it are not known.
The equation was already treated in the RG method by one of the
   present authors (TK), 
   and an approximate solution was constructed explicitly
 around the non-trivial fixed point\cite{kuni97}.
A numerical comparison of the results with the exact solution
 shows that the solution given by the RG method well
  approximates the 
 exact solution in a global domain even when the small parameter 
 in the equation  $\eps, \eps'$ (see below) are as large as $0.8$
\cite{numeric}.
The purpose of the present subsection is to apply the RG method 
for  analyzing the equation around the unstable  fixed point 
(the origin).
  
Lotka--Volterra equation reads
\begin{equation}\label{eqn:3-1}
\left\{
\begin{array}{ll}
\dot x = &ax-\epsilon xy\\
\dot y = &-by+{\epsilon}'xy\end{array}\right.
\end{equation}
Here,$x=x(t)$,$y=y(t)$ and $a,b,\epsilon,\epsilon'$ are positive
 constants.
In this work, we treat the case where $0<\eps<1$ and $0<\eps'<1$ so that the
perturbation theory can be applied.

There are two fixed points;
(i)\ $ x=y=0$
 and (ii)\ $x=b/\eps',\quad y=a/\eps$.
One can see that the fixed point (i) is asymptotically unstable.
An approximate but globally  valid solution around the second 
fixed point was obtained in \cite{kuni97}.
  In this paper, we treat the first fixed point.

With  the new variables defined by 
$x=\frac{\epsilon}{\epsilon'}\xi, \quad y=\eta$
(\ref{eqn:3-1}) is converted to 
\beq
\label{eq:lv_1}
\frac{d\bfu}{dt}=A\bfu+\eps \bfF (\bfu),
\eeq
where 
\beq
\bfu ={\xi\choose{\eta}},\quad  
A=\left(
\begin{array}{ll}
                 a & 0\\
                 0 & -b
\end{array}
   \right),
\quad  
\bfF(\bfu)=-\xi \eta {1\choose{-1}}.
\eeq
One see that the fixed point (i) is asymptotically unstable in the linear 
approximation.
Our aim here is to obtain the dynamics around the unstable fixed point
 and construct the unstable manifold using the RG method.
  
For later convenience, we introduce the normalized eigenvectors 
 of $A$;
\beq
A\bfU_1=a\bfU_1, \quad A\bfU_2=-b\bfU_2,
\eeq
where
$\bfU_1=\ ^t(1, 0)$ and 
$\bfU_2=\ ^t(0, 1)$.

Now let us apply the RG method to construct 
an approximate solution valid in a global domain.
We suppose that the equation is solved up to arbitrary $t=t_0$ 
from the time, say  $t=0$, at which
 the genuine initial condition is imposed.
With this up-to-dated initial value $\bfW(t_0)$, 
 we try to construct the solution  to
 (\ref{eq:lv_1}) around  $t\sim t_0$
 by the perturbation theory, expanding $\bfu$ as
$\bfu=\bfu_0+\epsilon\bfu_1+\epsilon^2\bfu_2+0(\epsilon^2)$.
Thus $\bfu$ may be written as 
$\bfu =\bfu(t; t_0)$.
We also expand the initial value  
$\bfW(t_0)=\bfW_0 +\eps \bfW_1+\eps ^2\bfW_2 + o(\eps^3)$.

The equations to be solved are 
\begin{eqnarray}
(\frac d{dt}-A)\bfu_0&=&0, \label{eqn:3-15}\\
(\frac d{dt}-A)\bfu_1&=&-\xi_0\eta_0(\bfU_1-\bfU_2), \label{eqn:3-16}\\
(\frac d{dt}-A)\bfu_2&=&-(\xi_0\eta_1+\xi_1\eta_0)
                         (\bfU_1-\bfU_2), \label{eqn:3-17}
\end{eqnarray}
and so on, where $\ ^t(\xi_i, \eta_i)=\bfu _i$ ($i=0, 1, 2 ...$).
It turns out that one needs to treat separately depending on whether
 $a\not= b$ or $a=b$. 
\\

\underline{Case A: $a=b$}\\ 
This is the interesting  case where secular terms appear and the RG
 method plays a role to construct an approximate solution in a global 
 domain.
The unperturbed solution reads
\begin{eqnarray}
{\bfu}_0 = C_1(t_0)e^{at}\bfU_1+C_2(t_0)e^{-at}\bfU_2,\label{eqn3-23}
\eeq
 implying the initial condition
$\bfW _0(t_0)=C_1(t_0)e^{at_0}\bfU_1+C_2(t_0)e^{-at_0}\bfU_2$.
The first order solution now reads
\beq
{\bfu}_1 &=&
-C_1C_2\frac{1}{\d_t-A}(\bfU_1-\bfU_2)= \frac{C_1C_2}{a}(\bfU_1+\bfU_2),
  \nonumber \\ 
 \  &=&\bfW_1(t_0),\label{eqn3-24}
\eeq
which is constant and independent of $\bfu_0(t)$.

Using the formulae in Appendix A, 
the second order solution is also  obtained.
%
%
Collecting all the terms thus obtained, we have an approximate solution
 valid around $t\sim t_0$;
\beq
\bfu (t:t_0)&=&
      C_1(t_0)e^{at}\bfU_1+C_2(t_0)e^{-at}\bfU_2  
     +     \eps \frac{C_1C_2}{a}(\bfU_1+\bfU_2)
  \nonumber \\
 \ \  &+& \eps^2[
      -\frac{C_1^2C_2}{a}(t-t_0)e^{at}+\frac{C_1C_2^2}{2a^2}e^{-at}\}\bfU_1
\nonumber\\
       &+&
      \{\frac{C_1^2C_2}{2a^2}e^{at}+\frac{C_1C_2^2}{a}(t-t_0)e^{-at}\}\bfU_2
      ]
       \label{eq:lv-final}.
\eeq           
Applying the RG equation $\d \bfu/\d t_0\vert _{t_0=t}=0$ to 
(\ref{eq:lv-final}),
 we have
\begin{equation}\label{eqn:3-29}
\dot{C_1}+\epsilon^2\frac{C_1^2C_2}{a}=0,
\quad
\dot{C_2}-\epsilon^2\frac{C_1C_2^2}{a}=0 .
\end{equation}
Noting the constraint $C_1C_2=\mbox{const.}=c$ given by (\ref{eqn:3-29}),
 we reach 
\begin{equation}\label{eqn:3-32}
C_1(t)=c_1e^{-\frac{\epsilon^2c}{a}t},
\quad
C_2(t)=\frac {c}{c_1}e^{\frac{\epsilon^2c}{a}t},
\end{equation}
with $c_1$ being a constant.

Our solution is given by the initial value $\bfW(t)$;
\beq
\label{eq:3-33}
\bfu(t; t)=\bfW(t)&=&
C_1(t)e^{at}\bfU_1+C_2(t)e^{-at}\bfU_2+
 \eps \frac{c}{a}(\bfU_1+\bfU_2)\nonumber \\ 
 \ \ &+& \eps^2 c\{\frac{C_2(t)}{2a^2}e^{-at}\bfU_1
       + \frac{C_1(t)}{2a^2}e^{at}\bfU_2
                 \},
\eeq
 with  $C_{1,2}(t)$ given by (\ref{eqn:3-32}).
Introducing $c_2$ by $c=c_1c_2$, the respective components are given by
\begin{eqnarray}
\xi(t;t)
&=&
c_1e^{(a-\frac{\epsilon^2c_1c_2}{a})t}+\frac{\epsilon c_1c_2}{a}
+
\frac{\epsilon^2c_1c_2^2}{2a^2}e^{-(a-\frac{\epsilon^2c_1c_2}{a})t}
+
o(\epsilon^2),\label{eqn:3-35}\\
\nonumber\\
\eta(t;t)
&=&
c_2e^{-(a-\frac{\epsilon^2c_1c_2}{a})t}+\frac{\epsilon c_1c_2}{a}
+
\frac{\epsilon^2c_1^2c_2}{2a^2}e^{(a-\frac{\epsilon^2c_1c_2}{a})t}
+
o(\epsilon^2).\label{eqn:3-36}
\end{eqnarray}
Here, the constants $c_1$ and $c_2$ are determined by the initial 
condition imposed at $t=0$.

Some remarks are in order:
(1)\  Our solution (\ref{eq:3-33}) shows that the  speed to approach to 
and to escape from the origin 
 is shifted from $a$ to $a-\eps ^2c_1c_2/a\equiv \alpha$, 
 which is dependent on 
 the initial condition as expressed by $c_1$ and $c_2$.
\ (2)\  The unstable manifold can be constructed explicitly from 
    (\ref{eqn:3-35}) and (\ref{eqn:3-36}).  Solving $\e^{\alpha t}$
     and $\e^{-\alpha t}$ from these equations and making the product of
     them, we have
\beq
\label{eq:lv-man}
c=(X-\theta Y)(Y-\theta X),
\eeq
where
$X=\frac{\eps'}{\eps}x -x_0,\ \ Y=y-y_0, \ \ $
$ \theta =\frac{\eps^2c}{2a^2}\ \ {\rm and}\ \  $
$x_0=y_0=\eps {c \over a}$.
(\ref{eq:lv-man}) shows that the unstable manifold is the hyperbolic curve
 which has the non-orthogonal asymptotic lines coming out of the point 
$\ ^t(x_0, x_0)$ with the slopes $\theta$ and $1/\theta$, respectively .


\underline{Case B: $a\ne b$}\\ 
This is a rather trivial case because no secular terms appear, and the
RG method does not play any role.
Therefore we only write down the result:
\begin{eqnarray}
\xi(t)&=&
C_1e^{at}+\epsilon\frac{C_1C_2}{b}e^{(a-b)t}\nonumber\\
&+&
\epsilon^2\{\frac{C_1C_2^2}{2b^2}e^{(a-2b)t}-
\frac{C_1^2C_2}{a(a-b)}e^{(2a-b)t}\}
+ 0(\epsilon^2),\label{eqn:3-21}\\
\nonumber\\
\eta(t)&=&
C_2e^{-bt}+\epsilon\frac{C_1C_2}{a}e^{(a-b)t}\nonumber\\
&+&
\epsilon^2\{\frac{C_1^2C_2}{2a^2}e^{(2a-b)t}+
\frac{C_1C_2^2}{b(a-b)}e^{(a-2b)t}\}
+ 0(\epsilon^2)\label{eqn:3-22},
\end{eqnarray}
where $C_1$ and $C_2$ are constant.
\newpage
\setcounter{equation}{0}
\section{Applications II:\ Pulse interactions}

In this section, we apply the RG method to obtain 
the dynamics of interacting pulses (or fronts) in one dimension.
Some yeas ago, a systematic method was developed by Ei and Ohta\cite{eo}
 to this 
problem on the basis of the phase dynamics approach \cite{dissipative2}
which involves a solvability condition.
We shall show how the dynamics of interacting pulses obtained by them 
 can be derived mechanically in the present method virtually without any
 assumptions.
 In other words, we shall derive the phase equations describing
  the front and pulse interactions in the RG method for the first time, 
although interface dynamics in spinodal decomposition\cite{phase},
  a diffusion equation\cite{sasa}
 and Kuramoto-Sivashinski equation \cite{maruo} as the phase 
  equations have been derived 
in the RG method by others.
As representative examples of pulse dynamics, we take up the kink-anti-kink 
interactions in the time-dependent Ginzburg-Landau (TDGL) equation and 
the soliton-soliton interaction in the Kortweg-de Vries (KdV) equation.

\subsection{Kink-anti-kink interaction in TDGL equation}

The TDGL equation we study is given by 
\beq
\label{eq:tdgl-1}
\frac{\d u}{\d t}=\eps^2 \frac{\d^2 u}{\d x^2}+f(u),
\eeq
with
\beq
f(u)={1\over 2}u(1-u^2),
\eeq
where $u=u(x, t)$ is a real scalar function and $0<\eps < 1$.
We remark that (\ref{eq:tdgl-1}) has a stationary solution
 
\beq
u(x,t)=\pm \tanh {{x-h}\over {2\eps}}\equiv \pm U(x-h),
\eeq
where $U(x)$ satisfies $U(\pm \infty)=\pm 1$ and $U(0)=0$.
$U(x-h)$ $(-U(x-h))$ is called a kink (an anti-kink) at the position $x=h$.

\subsubsection{Interaction of one kink and anti-kink}

We first consider the interaction of one kink and anti-kink.
Suppose the initial data at $t=0$ are given by
\[
U_0(x; x_1(0), x_2(0)) = U(x - x_1(0)) + \{-U(x - x_2(0)) - 1\} ,
\]
where $x_2(0) - x_1(0) >> \ep$; i.e., the kink (anti-kink) is located at 
$x=x_1(0)$ ($x=x_2(0)$).
Since $U_0(x; x_1(0), x_2(0))$ is not a solution to Eq.(\ref{eq:tdgl-1}),
 the position of the kink $x_1$ and the anti-kink $x_2$ will move 
 slowly. Our task is to find the equation governing the dynamics of 
 $x_1(t)$ and $x_2(t)$ at $t>0$.
 
To apply the RG method to solve this problem, we 
first identify the small parameter  and the integral 
constants in the unperturbed solution, which will move by the 
perturbation.
 When $x_2-x_1>>\eps$,
$U_0(x; x_1, x_2)$ in the neighborhood of $x = x_1$ may be represented by
\[
U_0(x; x_1, x_2) = u_0(t;x_1)(x) + \delta s(x - x_2),
\quad u_0(t; x_1)(x) = U(x - x_1),
\]
where $\delta $ is a small parameter because the effect of
the anti-kink $-U(x - x_2) - 1$ is small around $x = x_1$;
 the order of $\delta$ is ${\rm exp}(-(x_2 - x_1)/\eps)$.
This implies that as the unperturbed solution around $x\sim x_1$,
we may take $u_0(t;x_1)$, where $x_1$ is a constant.
Thus we are led to represent the solution at $t\sim \forall t_0$ by 
\beq
\label{eq:tdgl-2-1}
u(x,t)=u_0(t; x_1(t_0))(x)+\delta s(x - x_2(t_0))+v(x, t),
\eeq
with $v(x,t)$ as small as $\delta$. 
Inserting (\ref{eq:tdgl-2-1}) into (\ref{eq:tdgl-1}), we have 
\begin{eqnarray}
\label{eq:emy31-1}
\d_tv=
Av + (f'(u_0)-f'(1))\delta s
+ O(\delta ^2 + \vert v\vert ^2),
\eeq
with
\beq
Av\equiv \biggl(\eps^2\frac{\d^2}{\d x^2}+f'(u_0)\biggl)v.
\eeq
Here, we have made use of the fact that $u_0$ and $\delta s+1$ are 
stationary solutions to Eq.(\ref{eq:tdgl-1}); we also note
 that $f'(1) = f'(-1)$.

\rf{eq:emy31-1} is an equation of the type discussed in \S\S 4.1.
Indeed, owing to the translational invariance of the TDGL equation 
(\ref{eq:tdgl-1}), the 
self-adjoint operator $A$ has a zero
eigenvalue with the eigenfunction $U_1$ together with the corresponding
adjoint eigenfunction $\widetilde{U}_1$
\beq
U_1 (x) = \widetilde{U_1}(x) = \pa _xU(x - x_1).
\eeq
Therefore, one can obtain the approximate solutions of \rf{eq:emy31-1}
and the dynamics of $x_1(t)$
 according to the procedure developed in \S\S 4.1.

 The solution of the 
eigenvalue problem for the operator $A$ may be seen in text books 
on quantum mechanics\cite{LL}.
Let $P$ be the projection operator onto the kernel of $A$
and $Q$ the one onto the subspace compliment to the kernel $A$;
we call the respective subspaces $P$- and $Q$-space. We notice that 
\beq
Pu=\frac{<U_1, u>}{<U_1, U_1>}U_1(x),
\eeq
where $<u,v>$ denotes the inner product defined by 
\beq
<u,v>=\int_{-\infty}^{\infty}u(x)v(x)dx.
\eeq

Now let us apply the RG method to solve $v$, thereby obtain the dynamics of
 $x_1(t)$. First we expand $v$ as
\beq
v=\delta v_1 + O(\delta ^2).
\eeq
The equation for $v_1$ read 
\[
\d _tv_1 = Av_1 + (f'(u_0)-f'(1))s .
\]
In this case, $W_0(t_0)$ in \S\S 4.1 is $U(x - x_1(t_0))$.
$W_1(t_0)$ in $Q$-space is given by $(-A)^{-1}Q\{ f'(u_0) - f'(1)\} s$
by \rf{eq:init-1} and $v_1$ is

\beq
v_1(t;t_0)(x) = (t-t_0)P(f'(u_0)-f'(1))s + A^{-1}Q(f'(u_0)-f'(1))s.
\eeq
Thus, we have an approximate function
\[
u(t;t_0)(x) 
= U(x - x_1(t_0)) + \delta v_1(t;t_0)(x) + \delta s(x - x_2(t_0)).
\]
Applying the RG equation  to $u(t;t_0)(x)$,
one has 
\beq
0&=&\frac{\d u}{\d t_0}\biggl \vert_{t_0=t}=
-\dot{x}_1\d _xU(x-x_1)-P(f'(u_0)-f'(1))\delta s(x - x_2) + O(\delta ^2),
 \nonumber  
\eeq
leading to the dynamics governing $x_1$,
\begin{eqnarray}
\nonumber
\dot{x}_1 &=&
-\frac{<U_1,(f'(u_0)-f'(1))\delta s >}{<U_1, U_1>}
+ O(\delta ^2) \\
\nonumber
&=&
-\frac{<\d _xU(x - x_1),(f'(U(x-x_1)) - f'(1))
\{ -U(x-x_2) - 1\} >}{<U_1, U_1>} + O(\delta ^2) \\
\nonumber
&=&
-\frac{<\d _xU(x),(f'(U(x)) - f'(1))
\{ -U(x-(x_2 - x_1)) - 1\} >}{<U_1, U_1>} + O(\delta ^2) \\
&=&
12\eps e^{-\frac{x_2 - x_1}{\eps}} + O(\delta ^2),
\eeq
which coincides with the result by Carr-Pego\cite{cp},
Fuco-Hale\cite{FH} and Ei-Ohta\cite{eo} where an explicit evaluation  of 
the inner products is also given.

\subsubsection{Kink-anti-kink interaction in  the presence of infinite
kinks and anti-kinks}

We next consider an initial value problem of Eq.(\ref{eq:tdgl-1}) 
with the initial condition where infinite kinks (anti-kinks) are located 
periodically at $x=h+2n$ \ ($x=-h+2n$) with $n=0, \pm 1, \pm 2...$
\cite{cp,FH}:
We suppose that the intervals of a kink and the neighboring 
anti-kinks are much larger than the width of a kink (anti-kink), i.e.,
$h\gg \eps$.  In this situation,
the problem may be formulated as an initial value problem with one kink at 
$x=h$ in a finite domain $0<x<1$ with a Neumann boundary 
condition, i.e.,
\beq
\frac{\d u}{\d x}\biggl\vert_{x=0}=
\frac{\d u}{\d x}\biggl\vert_{x=1}.
\eeq

The initial profile in this case can be approximately represented 
by the function
\beq
U_0(x; h)&=&U(x-h)+\{ -U(x+h-2)-1\}+\{ -U(x+h)+1\}, \\ 
 \ \        &\equiv& u_0(x)+\delta _1r(x + h - 2) + \delta _2 \ell (x + h)
\eeq
where the second and the third terms denote the small effects coming 
from the anti-kink at $x=2-h$ and $x=-h$, respectively;
the smallness of the effects are represented by the parameters 
$\delta _1$ and $\delta _2$, 
which are the orders of ${\rm exp}(-2h/\eps)$ and ${\rm exp}(-2(1-h)/\eps)$,
respectively.

Let us represent the solution for $t>0$ by 
\beq
\label{eq:tdgl-2}
u(x,t)=U_0(x; h)+v(x, t),
\eeq
where $\vert v(x, t)\vert $ is supposed to be as small as $\delta _1$ 
and $\delta _2$. 
Inserting (\ref{eq:tdgl-2}) into Eq.(\ref{eq:tdgl-1}), we have
\begin{eqnarray}
\label{eq:emy31}
\d_tv
=
Av + (f'(u_0)-f'(1))(\delta _1r + \delta _2 \ell ) 
+ O(\delta _1^2+\delta _2^2 + \vert v\vert ^2)
\eeq
in  a similar manner with  \rf{eq:emy31-1}.
Here, $Av\equiv ( \eps^2\frac{\d^2}{\d x^2}+f'(u_0))v$
and the eigenfunction associated with zero eigenvalue in this case
is $U_1 = \d _xU(x - h)$.

Let us apply the RG method to solve $v$, thereby obtain the dynamics of
 $h(t)$. First we expand $v$ as
\beq
v=\delta _1v_{1,0}+\delta _2v_{0,1}+ O(\delta _1^2 + \delta _2^2).
\eeq
The equations for $v_{1,0}$ and $v_{0,1}$ read 
\beq
\d _tv_{1,0}=Av_{1,0}+(f'(u_0)-f'(1))r, \\ 
\d _tv_{0,1}=Av_{0,1}+(f'(u_0)-f'(1))\ell,
\eeq
respectively.
These are of the same form as treated in \S\S 4.1, hence readily solved
to be
\beq
v_{1,0}(t;t_0)(x) = (t-t_0)P(f'(u_0)-f'(1))r
+ A^{-1}Q(f'(u_0)-f'(1))r,
\eeq
where $r = r(x + h - 2)$ and $h = h(t_0)$,
by choosing the initial value
\beq
v_{1,0}(t_0;t_0)(x) = -A^{-1}Q(f'(u_0)-f'(1))r(x + h(t_0) -2),
\eeq
and $v_{0,1}(x, t)$ with $r \rightarrow \ell $.
Thus, we have
\[
u(t;t_0)(x) = U(x - h(t_0)) + \delta _1v_{1,0}(t;t_0)(x)
+ \delta _2v_{0,1}(t;t_0)(x) + \delta _1r + \delta _2\ell .
\]
Applying the RG equation to $u(t;t_0)(x)$,
one has 
\beq
0 = \frac{\d u}{\d t_0}\biggl \vert_{t_0=t} &=&
-\dot{h}U'(x-h)-P(f'(u_0)-f'(1))(\delta_1 r(x + h -2) + 
\delta_2 \ell(x + h))\nonumber \\ 
  & & \; \; \;  + O(\delta _1^2 + \delta _2^2),
\eeq
leading to the dynamics governing $h$,
\begin{eqnarray}
\nonumber
\dot{h} &=&
-\eps\frac{<U_1,(f'(u_0)-f'(1))(\delta _1r+\delta _2\ell)>}
             {<U_1, U_1>} + O(\delta _1^2 + \delta _2^2) \\
&=&
-12\eps \left( e^{-\frac{2h}{\eps}} - e^{-\frac{2(1-h)}{\eps}} \right) 
+ O(\delta _1^2 + \delta _2^2) ,
\eeq
which also coincides with the known result by Carr-Pego\cite{cp} and 
Fusco-Hale\cite{FH}.

\subsection{Soliton-soliton interaction in KdV equation}

The KdV equation reads
\beq
\d_tu+6u\d_x u+\d_x^3u=0,
\eeq
which has a one-pulse solution given by 
\beq
u(x,t)=\frac{c}{2}{\rm sech}^2\biggl[\frac{\sqrt{c}(x-ct)}{2}\biggl]
 \equiv \varphi(x-ct; c),
\eeq
with $c$ being a velocity. 

We consider the following problem; When two sufficiently separated pulses
 with almost the
same velocities are located at 
$x_1(0)$ and $x_2(0)$ where $\vert x_2(0)-x_1(0)\vert\gg 1/\sqrt{c}$,
 how will the locations $x_i(t)$ ($i=1,2$) change at $t>0$?
To solve this problem, Ei and Ohta \cite{eo} started with the ansatz
\beq
u(x,t)=\varphi(x-ct-x_1; c+\dot{x}_1)+
\varphi(x-ct-x_2; c+\dot{x}_2)+b(x-ct,t).
\eeq
In the present work, we shall apply the RG method without any ansatz
to this problem and show that the same evolution equation as that obtained
 by Ei and Ohta is derived, thereby give a foundation of their 
 treatment.

To study the problem, it is convenient to change 
the independent variables to 
\beq
t=t,\quad z=x-ct,
\eeq
namely to change 
to the co-moving frame with the pulse. 
Then the equation is converted to 
\beq
\label{kdv;0}
\d_tu+F[u]=0,
\eeq
with
\beq
F[u]=-c\d_zu+6u\d_zu+\d_z^3u.
\eeq
We remark that
\beq
\label{kdv;id}
F[\varphi(z-b;c)]=0,
\eeq
with $b$ being an arbitrary constant.

Let us suppose that the solution around   $t=t_0>0$ is given 
by 
\beq
\label{kdv;1}
u(z,t) &=& \varphi(z-z_1(t_0);c)+\varphi(z-z_2(t_0);c)+v(z,t), \\ 
       &\equiv & \varphi^{(1)} + \varphi^{(2)} + v,
\eeq
where $\vert z_2-z_1\vert$ is sufficiently large.
To study the effect coming from the other pulse, it is sufficient to 
consider the case of either $z\sim z_1$ or $z\sim z_2$.
Then $v$ is considered to be small.

Substituting Eq.(\ref{kdv;1}) into (\ref{kdv;0}), we have
the equation governing $v$ as
\beq
\label{eq:emy32-1}
\d_tv+F'[\varphi^{(1)} + \varphi^{(2)}]v 
+ 6\d_z(\varphi^{(1)}\varphi^{(2)}) + O(\vert v\vert^2)
=0.
\eeq
Here we have used the identity 
$F[\varphi^{(1)} + \varphi^{(2)}]
= F[\varphi^{(1)} + \varphi^{(2)}] - F[\varphi^{(1)}] - F[\varphi^{(2)}]
  =6\d_z(\varphi^{(1)}\varphi^{(2)})$ on account of Eq.(\ref{kdv;id}).

Now let $z\sim z_1$, then $\varphi^{(2)}$ is small and we may put 
\beq
\varphi^{(2)} = \varphi (z - z_2(t_0)) = \delta g(z - z_2(t_0)),
\eeq
with a small parameter $\delta$ whose order is $e^{-\sqrt{c}(z_2-z_1)}$.
Then, \rf{eq:emy32-1} of $v$ becomes
\begin{equation} \label{eq:emy32-2}
\d_tv = A^{(1)}v - 6\delta \d_z(\varphi^{(1)}g ) + O(\delta ^2 + |v|^2),
\end{equation}
where 
\[
A^{(1)} = -F'[\varphi^{(1)}]
= c\d_z - \d_z^3 - 6(\d_z\varphi^{(1)} + \varphi^{(1)}\d_z).
\]
Transforming $z - z_1(t_0)$ to $z'$, we see \rf{eq:emy32-2} becomes
\begin{equation} \label{eq:emy32}
\pa _tv = Av - \delta 6\pa _{z'}(\varphi g^{(h)}) + O(\delta ^2 + |v|^2),
\end{equation}
where $A = -F'(\varphi )$ and $\delta g^{(h)}(z') = \varphi (z' - h)$
with $h = z_2(t_0) - z_1(t_0)$. 

\rf{eq:emy32} is an equation of the
type discussed in \S\S 4.3;
the linear operator $A$ has a Jordan cell reflecting the translational
invariance of KdV equation and the arbitrariness of the velocity 
 of a pulse,
\beq
AU_1 = 0,\quad AU_2 = U_1 ,
\eeq
where
\beq
U_1 = \d_z\varphi , \quad U_2 = -\d_c\varphi .
\eeq
The adjoint operator $A^{\dagger} $ of $A$ reads
\beq
A^{\dagger} = -c\d_{z'} + \d_{z'}^3 + 6\varphi \d_{z'} ,
\eeq
which has also a two-dimensional Jordan cell.
The zero mode $\widetilde{U}_2$ 
of $A^{\dagger}$ is found to be \cite{eo}
\beq
\widetilde{U}_2(z') = \varphi (z';c)
\eeq
while these eigenfunctions have not been normalized as \rf{eq:ein1} yet.
It is known \cite{PW}
 that there exists a function $\widetilde{U}_1$ which satisfies 
\beq \label{eq:wtu-2}
A^{\dagger}\widetilde{U}_1 = \widetilde{U}_2.
\eeq
The explicit form of it will be given later.

Let $P$ be the projection operator onto the subspace spanned by 
$U_1$ and $U_2$, and $Q$ onto the subspace compliment of the
$P$-space.

Noting that $\vert v\vert \le O(\delta)$ and
expanding $v$ as
\beq
v=\delta v_1+\delta^2v_2+\cdots,
\eeq
one has for $v_1$ from \rf{eq:emy32}
\beq
\label{eq:kdv-pert}
\d_tv_1 = Av_1 - 6\d_{z'}(\varphi g^{(h)} ).
\eeq

Now we can proceed according to the general procedure given in
\S\S 4.3 as follows:
Note that $u_0$ and hence  $W_0(t_0)$ in \S\S 4.3   
correspond to $\varphi (z - z_1(t_0);c)$ because both of them comes from
the translation invariance with respect to $z$.
Let us solve the equation (\ref{eq:kdv-pert}) 
with the initial value $W_1(t_0)$. Then, we have
\beq
v_1(t;t_0)=
\e^{(t-t_0)A}W_1(t_0)-6\int_{t_0}^t\e^{(t-s)A}\d_{z'}(\varphi g^{(h)})ds,
\eeq
with 
\beq
W_1(t_0)=C(t_0)U_2 + QW_1(t_0)
\eeq
as discussed in \S\S 4.3.
Here we note that although $W_1(t_0)$ could have a $U_1$ component, 
its effect can be taken into account by 
a redefinition of $z_1(t_0)$ which is not yet determined.
According to the discussion given in \S\S4.3, 
the $Q$-component of $W_1$ should be
\beq
QW_1(t_0) = -A^{-1}Q(-6\d_{z'}(\varphi g^{(h)})).
\eeq
Thus we have
\beq
v_1(t;t_0)(z')
&=& \left\{
(t-t_0)C(t_0) + (t-t_0)\alpha + \frac{1}{2}(t-t_0)^2\beta \right\}U_1
\nonumber \\
&& + \{ C(t_0) + (t-t_0)\beta \}U_2
- A^{-1}Q(-6\d_{z'}(\varphi g^{(h)}))
\eeq
and the approximate function
\beq
u(t;t_0)(z) = \varphi(z - z_1(t_0);c) + \delta v_1(t;t_0)(z - z_1(t_0))
+ \delta g(z - z_2(t_0)),
\eeq
where
\beq
\alpha=\frac{<\widetilde{U}_1, -6\d_{z'}(\varphi g^{(h)})>}
{<\widetilde{U}_1,U_1 >},
\quad 
\beta=\frac{<\widetilde{U}_2, -6\d_{z'}(\varphi g^{(h)})>}
{<\widetilde{U}_2, U_2 >}.
\eeq

Applying the RG equation to $u(t;t_0)(z)$,
one has
\beq \label{eq:emy33}
\frac{\d u}{\d t_0}\biggl\vert_{t_0=t}
=
-\dot{z}_1\d_z\varphi^{(1)} + \delta\left[
\{ -C - \alpha \}U_1 ^{(1)} + \{ \dot{C} - \beta \}U_2 ^{(1)}
\right] + O(\delta^2),
\eeq
where $U_1 ^{(1)}(z) = U_1 (z - z_1(t_0)) = \pa _z\varphi (z - z_1(t_0))$
and $U_2 ^{(1)}$ is similarly given. Since $U_1 ^{(1)} 
= \pa _z\varphi ^{(1)}$,
\rf{eq:emy33} leads to 
\beq
\dot{z}_1=-\delta C(t) -\delta \alpha, \quad 
\dot{C}=\beta.
\eeq
Recalling that       
\[
\delta \alpha = \tilde{\alpha} =
\frac{<\widetilde{U}_1, -6\d_{z'}(\varphi \delta g^{(h)})>}
                 {<\widetilde{U}_1, U_1 >}
= \frac{<\widetilde{U}_1, -6\d_{z'}(\varphi \varphi^{(h)})>}
                 {<\widetilde{U}_1, U_1 >}
\]
and 
\[
\delta \beta = \tilde{\beta} =
\frac{<\widetilde{U}_2, -6\d_{z'}(\varphi \delta g^{(h)})>}
{<\widetilde{U}_2, U_2 >} 
= \frac{<\widetilde{U}_2, -6\d_{z'}(\varphi \varphi^{(h)})>}
{<\widetilde{U}_2, U_2 >} ,
\]
where $\varphi^{(h)} (z') = \varphi (z' - h)$,
and putting $\delta C(t)=C_1(t)$, we finally obtain 
 the equations of motion
 governing the position of the kink and the speed of it,
\beq
\label{eq:emy34}
\dot {z_1} = -C_1 + \frac{<\widetilde{U}_1, -6\d_{z'}(\varphi \varphi^{(h)})
>}
                 {<\widetilde{U}_1 ,U_1 >} ,
\quad 
\dot{C_1} = \frac{<\widetilde{U}_2, -6\d_{z'}(\varphi \varphi^{(h)})>}
{<\widetilde{U}_2, U_2 >}.
\eeq
Similarly, one can readily obtain the equation for $\dot{z_2}$.
This is the main result in this subsection.

Eliminating $C_1$ from the above equations, one can compare our result 
with those given by Ei and Ohta\cite{eo}.
We first notice 
that 
\beq
\label{eq:order}
\frac{d}{dt}\tilde{\alpha} =
\frac{d}{dt}\frac{<\widetilde{U}_1, -6\d_{z'}(\varphi \varphi^{(h)})>}
                 {<\widetilde{U}_1, U_1 >}
= O(\delta ^2),
\eeq
because $\tilde{\alpha}$ depends only on $h = z_2 - z_1$ and 
$\dot{h} = O(\delta)$.
Then we have 
\beq
\label{eq:emy35}
\ddot{z}_1= \frac{<\widetilde{U}_2, 6\d_{z'}(\varphi \varphi^{(h)})>}
{<\widetilde{U}_2, U_2 >}
= -16c^{5/2}e^{-\sqrt{c}(z_2 - z_1)} + O(\delta ^2) ,
\eeq
and similarly
\beq
\label{eq:emy36}
\ddot{z}_2 \simeq 16c^{5/2}e^{-\sqrt{c}(z_2 - z_1)} ,
\eeq
which coincide with the result by Ei and Ohta\cite{eo}.
In short, we have derived the equation describing the 
soliton-soliton dynamics in the KdV equation given by Ei and Ohta
 with fewer ansatz on the basis of the 
RG method.

As noticed above, the system \rf{eq:emy34}  have 
more information for the soliton-soliton dynamics 
than \rf{eq:emy35} (or \rf{eq:emy36})
because \rf{eq:emy34} gives the 
the change of the velocity $C_1$ as well as the speed of the position $z_1$.
It is intriguing  to see an explicit form of \rf{eq:emy34}, which 
will provide more detailed information of the dynamics of the solitons.

The right hand side of the equation for $\dot{C}_1$ 
in \rf{eq:emy34} is already 
given in \rf{eq:emy35}. Let us calculate the r.h.s. of
$\dot{z}_1$.
%
 The adjoint eigenfunction $\widetilde{U}_1$ is given by 
\[
\widetilde{U}_1 = \dint ^z_{-\infty}\pa _c\varphi (s;c)ds + \theta \varphi,
\]
where $\theta = -M_1/M_2$ with $M_1=$
$1/2\cdot \left( \dint ^{\infty}_{-\infty}\pa _c\varphi ds\right) ^2$
and $M_2 = <\varphi , \pa _c\varphi >$\cite{PW}.
Note that $\la \tilde{\bfU}_1, \bfU _2 \ra = 0$ holds.
 $M_2$ is evaluated  to be $\sqrt{c}/{2}$\cite{eo}.
We here evaluate $M_1$.
We first notice that $\pa _c\varphi = \varphi/c+ z\pa _z\varphi/2c$.
Then by a partial derivative, we have 
\begin{eqnarray*}
\dint ^z_{-\infty}\pa _c\varphi ds 
&=&
\frac{1}{2c} \left(\dint ^z_{-\infty}\varphi ds + z\varphi\right) \\
& \rightarrow &
\dfrac{1}{2c}\dint ^{\infty}_{-\infty}\varphi ds
= \dfrac{1}{\sqrt{c}} \; \quad (z \rightarrow \infty),
\end{eqnarray*}
hence  $M_1 =1/2c$ and 
\begin{equation} \label{eq:theta}
\theta = -1/c\sqrt{c}.
\end{equation}

Since $\widetilde{U}_1$ is related with $\widetilde{U}_2 (= \varphi)$
 through \rf{eq:wtu-2}, we see that 
\[
6\varphi \pa _z\TU _1 = c\pa _z\TU _1 - \pa _z^3\TU _1 + \varphi .
\]
Then performing a partial derivative, we have 
\begin{eqnarray}
\nonumber
\lng \TU _1, -6\pa _z(\varphi \varphi ^{(h)}) \rng 
&=&
\lng c\pa _z\TU _1 - \pa _z^3\TU _1 + \varphi , \varphi ^{(h)} \rng \\
& \simeq &
2ce^{-\sqrt{c}h} \dint ^{\infty}_{-\infty}
e^{\sqrt{c}z} \{ c\pa _z\TU _1 - \pa _z^3\TU _1 + \varphi \} dz,
\label{eq:emy38}
\end{eqnarray}
where we have used the relation 
 $\varphi ^{(h)}(z) \simeq 2ce^{\sqrt{c}(z - h)}$ for $h >> 1$.
The remaining integral can be performed using the 
the asymptotic forms of the derivatives of $\TU _1$,
\begin{equation} \label{eq:emy39}
\pa _z\TU _1 \simeq (4 - \sqrt{c}z)e^{-\sqrt{c}z},\;
\pa _z^2\TU _1 \simeq -\sqrt{c}(5 - \sqrt{c}z)e^{-\sqrt{c}z}.
\end{equation}
%
Thus, for $L >> 1$ we have
\begin{eqnarray}
\dint ^L_{-\infty}e^{\sqrt{c}z}\pa _z\TU _1 dz 
&=&
e^{\sqrt{c}L}\TU _1(L) - \sqrt{c}\dint ^L_{-\infty}e^{\sqrt{c}z} \TU _1 dz,
\label{eq:emy40} \\
\dint ^L_{-\infty}e^{\sqrt{c}z}\pa _z^3\TU _1 dz 
&=&
-\sqrt{c}(9 - 2\sqrt{c}L)
+ ce^{\sqrt{c}L} \TU _1(L)
- c\sqrt{c} \dint ^L_{-\infty}e^{\sqrt{c}z} \TU _1 dz .
\label{eq:emy41}
\end{eqnarray}

Finally, the third term of \rf{eq:emy38} is evaluated to be
;
\begin{eqnarray}
\nonumber
\dint ^L_{-\infty}e^{\sqrt{c}z}\varphi dz 
&=& \sqrt{c}\dint ^{L'}_{-\infty}
e^{2z} {\rm sech}^2z dz  \quad (L' = \sqrt{c}L/2),\\
\nonumber
&=&
2\sqrt{c}\left( 2L' + \log (1 + e^{-2L'}) 
+ \dfrac{1}{e^{2L'} + 1} - 1 \right), \\
& \simeq &
2\sqrt{c}(2L' - 1) = 2\sqrt{c}(\sqrt{c}L - 1) .
\label{eq:emy42}
\end{eqnarray}
Inserting \rf{eq:emy40} $\sim $ \rf{eq:emy42} into \rf{eq:emy38},
we have
\begin{eqnarray}
\nonumber
\lng \TU _1, -6\pa _z(\varphi \varphi ^{(h)}) \rng 
& \simeq &
2ce^{-\sqrt{c}h}
\lim _{L \rightarrow \infty}
\dint ^L_{-\infty}
e^{\sqrt{c}z} \{ c\pa _z\TU _1 - \pa _z^3 + \varphi \} dz \\
& = &
14\sqrt{c} . \label{eq:emy43}
\end{eqnarray}
The normalization integral is evaluated to be
\[
\lng \TU _1, U_1 \rng 
= - \lng \dint ^z_{-\infty}\pa _c\varphi dz, \pa _z\varphi \rng
=
-\lng \pa _c\varphi , \varphi \rng = -\sqrt{c}/2.
\]
Thus we finally obtain
\beq
\label{eq:final-1}
\frac{<\widetilde{U}_1, -6\d_z(\varphi \varphi^{(h)})>}
                 {<\widetilde{U}_1 ,U_1 >}
\simeq -28ce^{-\sqrt{c}h},
\eeq
and hence 
 the equations for $z_1$ and $C_1$ is
\beq \label{eq:ein2}
\left\{
\begin{array}{lcl}
\dot{z}_1 &=& -C_1 - 28ce^{-\sqrt{c}(z_2 - z_1)}, \\
\dot{C}_1 &=& 16c^{5/2}e^{-\sqrt{c}(z_2 - z_1)} ,
\end{array}
\right.
\eeq
together with a similar equation for $z_2$.
Notice that Eq.(\ref{eq:ein2}) is consistent with the order estimate of
 Eq.(\ref{eq:emy35}).
\newpage
\section{Brief summary and concluding remarks}

We have formulated the RG method as a powerful tool for reduction of evolution
equations in terms of 
 the notion of invariant manifolds, starting from the exact
Wilson  RG equation. We have given an argument as to why 
 $t_0$ should be set $t=0$ in the
 perturbative RG method.
We have shown that the perturbative RG method constructs
 invariant manifolds successively as the initial value of 
evolution equations; the integral constants in the unperturbative 
solution constitutes natural coordinates of the invariant manifold
when the linear operator $A$ in the evolution equation has no Jordan cell.
When $A$ has a Jordan cell, there is a slight complication because 
the dimension of the invariant manifold 
is to change by the perturbation.
The RG equation determines the slow motion
of the integral constants in the unperturbative solution
 on the invariant manifold.
 We have worked out several examples to demonstrate our formulation.
 We have emphasized that the underlying structure of the reduction by the
 RG method completely fits to the universal structure elucidated
 by Kuramoto\cite{kuramoto} a decade
 ago. The prescription suggested by the present formulation has turned out
  to be the same as that adopted in \cite{kuni95,kuni97}.
 We have applied the method to interface dynamics such as 
 kink-anti-kink  and soliton-soliton interactions in the latter of which 
 a linear operator having a  Jordan-cell structure appears.
 
In the present work, actual calculations are all based on the 
perturbation theory, although we have started the formulation 
with the exact flow equation. Recently, variants of the exact RG equations
or flow equations are applied to various problems such as 
the chiral symmetry
breaking in QCD\cite{rev_flow,jochen},
 the Bose-Einstein condensation in Alkali atoms \cite{strick} 
and so on. 
There, in stead of
 the perturbation theory, a truncation of the functional spaces are 
 usually employed as a practical method of calculations.
One may thus imagine that a possible practical method for 
applying the non-perturbative RG method could be a scheme similar to,
say, the Galerkin method\cite{galer}.
 Such a non-perturbative RG method should be 
most interesting for analyses of partial differential equations.
An attempt to apply a kind of non-perturbative RG equation to partial
differential equations is given in \cite{goldenfeld95}.

We have tried to formulate the RG method so that 
  the mathematical structure
 of the method becomes as transparent as possible.
 Then the present work could be a basis for 
 clarifying a possible relation between the RG method and
 another powerful theory for reduction of evolution equations
 called Whitham's averaging method\cite{whitham}:
 The latter method has been successfully used to extract the equations 
describing modulations of dispersive non-linear waves:  
Modulations of the phase function are given by slow variables which are
governed by so-called Whitham equations. One may thus imagine that 
the modulations as described by Whitham equations may also be given 
by the RG method as it gives the amplitude and phase equations.
 Or more strongly,
Whitham equations might be derived as RG ones\cite{takasaki}.
Furthermore, Whitham gave a foundation of his equations on the basis 
of a variational principle for the actions, which is reminiscent of the
fact that Wilsonian RG is formulated for effective actions which admit
a variational principle.

\vspace{1cm}

{\large {\bf Acknowledgement}} \\ 
We are grateful to the referee to have directed our attention to 
 Whitham's method.

\newpage
\renewcommand{\theequation}{A.\arabic{equation}}
\setcounter{equation}{0}
{\large {\bf Appendix A\ \ 
An efficient operator method of solution suitable for the RG method}}
\\

In this Appendix, we shall summarize  
rules for  obtaining appropriate special 
solutions of non-homogeneous equations appearing in the higher orders
in the perturbative RG method.
The essential point of the present method is to consider the equations 
with initial conditions at $t=\forall t_0$;
 the initial values are determined 
 so that the terms in the special solutions disappear 
 which either could be "renormalized away'' by a redefinition of the
 integral constants in the unperturbed solution or are the ones describing 
a rapid motion. We shall show that the simple rules to write 
down  the special solutions with such 
initial conditions can be summarized as an operator method.

Let us  consider the special solution 
to the   equation given by 
\beq
(\d_t-A)\bfu(t;t_0)=\bfF(t),
\eeq
with the initial condition at $t=t_0$,
\beq
\bfu(t_0, t_0)=\bfW(t_0).
\eeq
The solution reads
\beq
\bfu(t;t_0)=\e^{A(t-t_0)}\bfW(t_0)+
         \e^{At}\int_{t_0}^tds\e^{-As}\bfF(s).
\eeq
Let $A$ be a semi-simple matrix with 
 eigenvalues $\lambda_{\alpha}$\ ($\alpha=1, 2, ...$);
\beq
A\bfU_{\alpha}=\lambda_{\alpha}\bfU_{\alpha}.
\eeq
The case where  $A$ has a Jordan cell will be considered later.

Let 
\beq 
\bfF(t)=\e^{\lambda_{\alpha} t}\bfU_{\alpha}+
\e^{\lambda_{\beta} t}\bfU_{\alpha}+\e^{\lambda_{\alpha}t}\bfU_{\beta},
\eeq
with $\lambda_{\alpha}\not=\lambda_{\beta}$,
then the solution is evaluated to be
\beq
\bfu(t;t_0)&=&
\e^{A(t-t_0)}\left[\bfW(t_0)-\frac{1}{\lambda_{\beta}-A}\e^{\lambda_{\beta}t_0}
         \bfU_{\alpha}-
         \frac{1}{\lambda_{\alpha}-A}\e^{\lambda_{\alpha}t_0}
         \bfU_{\beta}\right]\nonumber \\ 
    & & 
    +(t-t_0)\e^{\lambda_{\alpha} t}\bfU_{\alpha}
     +\frac{1}{\lambda_{\beta}-A}\e^{\lambda_{\beta }t}\bfU_{\alpha}
     +\frac{1}{\lambda_{\alpha}-A}\e^{\lambda _{\alpha}t}\bfU_{\beta}.
\eeq
The first line suggests that the initial value should be chosen as 
\beq
\bfW(t_0)=\frac{1}{\lambda_{\beta}-A}\e^{\lambda_{\beta}t_0}
         \bfU_{\alpha}+
         \frac{1}{\lambda_{\alpha}-A}\e^{\lambda_{\alpha}t_0}
         \bfU_{\beta},
\eeq         
where the first term corresponds to the one 
which could be renormalized away by a redefinition of the unperturbed solution
 and the second term to a rapid motion. Thus we end up with the special solution
 given by 
\beq
\bfu(t;t_0)=(t-t_0)\e^{\lambda_{\alpha} t}\bfU_{\alpha}
     +\frac{1}{\lambda_{\beta}-A}\e^{\lambda_{\beta }t}\bfU_{\alpha}
     +\frac{1}{\lambda_{\alpha}-A}\e^{\lambda _{\alpha}t}\bfU_{\beta}.
\eeq     

\newcommand{\dtinv}{\frac{1}{\d_t-A}}

Then one may summarize the results as rules of an operator method
 for obtaining special solutions as follows;
\beq
\dtinv \e^{\lambda t}\bfU_{\alpha}
    &=&\frac{1}{\d_t-\lambda_{\alpha}}\e^{\lambda t}\bfU_{\alpha},
    \nonumber \\
    &=&\frac{1}{\lambda-\lambda_{\alpha}}
       \e^{\lambda t}\bfU_{\alpha}, \quad (\lambda\not=\lambda_{\alpha}),\\  
\ 
\dtinv \e^{\lambda _{\alpha}t}\bfU_{\alpha}
    &=&\frac{1}{\d_t-\lambda_{\alpha}}\e^{\lambda _{\alpha}t}\bfU_{\alpha}, 
    \nonumber \\ 
    &=&(t-t_0)\e^{\lambda _{\alpha}t}\bfU_{\alpha}.
\eeq
Similary, one can verify that 
\beq 
\dtinv (t-t_0)^n\e^{\lambda _{\alpha}t}\bfU_{\alpha}
    &=&\frac{1}{\d_t-\lambda_{\alpha}}(t-t_0)^n
       \e^{\lambda _{\alpha}t}\bfU_{\alpha},\nonumber \\ 
     &=&\frac{1}{n+1}(t-t_0)^{n+1}\e^{\lambda _{\alpha}t}\bfU_{\alpha}.
\eeq
Furthermore,when $\lambda\not=\lambda_{\alpha}$,
\beq
\dtinv (t-t_0)^n\e^{\lambda t}\bfU_{\alpha}
     &=&\e^{\lambda t_0}\frac{1}{\d_{\tau}-A}\tau^n\e^{\lambda \tau}
     \bfU_{\alpha}\biggl\vert_{\tau=t-t_0}
             ,\nonumber \\  
  &=&\e^{\lambda t_0}\d^n_{\lambda}\frac{1}{\d_{\tau}-A}\e^{\lambda \tau}
     \bfU_{\alpha}\biggl\vert_{\tau=t-t_0}.
\eeq
Hence, for example, 
\beq
\dtinv (t-t_0)\e^{\lambda t}\bfU_{\alpha}
     &=&\frac{1}{\lambda-A}\{(t-t_0)
       -\frac{1}{\lambda-A}\}\e^{\lambda t}\bfU_{\alpha},\\     
\dtinv (t-t_0)^2\e^{\lambda t}\bfU_{\alpha}
     &=&\frac{1}{\lambda-A}\{(t-t_0)^2
        -\frac{2}{\lambda-A}(t-t_0)
        +\frac{2}{(\lambda-A)^2}\}\e^{\lambda t}\bfU_{\alpha},
\eeq
where $A$ may be replaced with $\lambda_{\alpha}$.

Next, we 
consider the case where $A$ has a semi-simple zero eigenvalue;
\beq
A\bfU_0=0.
\eeq
Let P and Q be the projection operator onto the space spanned by $\bfU_0$
 and its orthogonal compliment, respectively.
If $\bfG$ is a constant vector, then one can easily verify that
\beq
\label{app;1}
\dtinv P\bfG&=&\frac{1}{\d_t}P\bfG=\int_{t_0}^tdsP\bfG=(t-t_0)P\bfG,\\ 
\label{app;2}
\dtinv Q\bfG&=&\frac{1}{-A}Q\bfG.
\eeq
Similarly,
\beq
\dtinv f(t)P\bfG&=&{1 \over {\d_t}}P\bfG=\int_{t_0}^tdsf(s)P\bfG,\\
\dtinv f(t)Q\bfG&=&{1\over {-A}}\sum_{n=0}^{\infty}(A^{-1}\d_t)^nQ\bfG=
-\sum_{n=0}^{\infty}f^{(n)}(t){1 \over {A^n}}Q\bfG,
\eeq
with $f^{(n)}(t)$ being the $n$-th derivative of $f(t)$.
Thus, for example,
\beq
\label{app;3}
\dtinv (t-t_0)^nP\bfG&=&{1 \over {n+1}}(t-t_0)^{n+1}P\bfG,\\ 
\label{app;4}
\dtinv (t-t_0)Q\bfG&=&-\biggl[(t-t_0){1\over A}+{1 \over {A^2}}\biggl]Q\bfG,\\ 
\label{app;5}
\dtinv (t-t_0)^2Q\bfG&=&-\biggl[(t-t_0)^2{1\over A}+2(t-t_0){1\over {A^2}}+
{1 \over {A^3}}\biggl]Q\bfG.
\eeq

Finally we consider the case where  $A$ has a two dimensional Jordan cell;
\beq
A\bfU_1=0,\quad A\bfU_2=\bfU_1.
\eeq
The adjoint has also a Jordan cell;
\beq
A^{\dag}\tilde{\bfU}_1=0,\quad A^{\dag}\tilde{\bfU}_2=\tilde{\bfU}_1.
\eeq
The adjoint operator $A^{\dag}$ is defined by 
$\la\bfV, A\bfU\ra=\la A^{\dag}\bfV, \bfU\ra$,
where $\la \bfV, \bfU\ra$ is the Hermitian inner product.
We define the projection operators $P$ and $Q$ onto the subspace 
$\{\bfU_1, \bfU_2\}$
 and its orthogonal compliment, respectively.
We suppose that the following normalization condition is satisfied;
\beq
\la\tilde{\bfU}_2, \bfU_1\ra=1,\quad \la\tilde{\bfU}_1, \bfU_2\ra=1.
\eeq
A vector $\bfU$ in the P-space is decomposed as 
$\bfU=\la\tilde{\bfU}_2, \bfU\ra\bfU_1+\la\tilde{\bfU}_1,\bfU\ra\bfU_2$.

Let $\bfG$ be a constant vector, then one has,
\beq
\dtinv f(t)P\bfG&=&{1\over {\d_t}}\sum_{n=0}^{\infty}(\d^{-1}_tA)^nf(t)P\bfG, 
        \nonumber \\ 
     &=&{1\over {\d_t}}\sum_{n=0}^{\infty}(\d^{-n}_tf(t))A^n
        [\la\tilde{\bfU}_2, \bfG\ra\bfU_1+\la\tilde{\bfU}_1, \bfG\ra\bfU_2],
        \nonumber \\ 
     &=&{1\over{\d_t}}[f(t)\{\la\tilde{\bfU}_2, \bfG\ra\bfU_1+
      \la\tilde{\bfU}_1, \bfG\ra\bfU_2\}+
        {1\over{\d_t}}f(t)\la\tilde{\bfU}_1, \bfG\ra\bfU_1],\nonumber \\
     &=& \int_{t_0}^tdsf(s)P\bfG+\int_{t_0}^tds\int_{t_0}^{s}ds'f(s')
     \la\tilde{\bfU}_1, \bfG\ra\bfU_1.
\eeq
Thus, for example,
\beq
\label{app;6}
\dtinv P\bfG&=& (t-t_0)P\bfG+{1\over 2}(t-t_0)^2\la\tilde{\bfU}_1,\bfG\ra
\bfU_1,\\
\dtinv (t-t_0)^nP\bfG&=& {1\over{(n+1)}}(t-t_0)^{n+1}P\bfG,\nonumber \\ 
          & &+{1\over {(n+2)(n+1)}}(t-t_0)^{n+2}
   \la\tilde{\bfU}_1, \bfG\ra\bfU_1.
\eeq
The formulae involving $Q\bfG$ are the same as those in the semi-simple case.

The extension to the case where $A$ has a higher dimensional Jordan cell is
easy. For instance, when $A$ has a three-dimensional Jordan cell such as
\beq
A\bfU_1=0,\quad A\bfU_2=\bfU_1,\quad A\bfU_3=\bfU_2,
\eeq
one can easily verify that
\beq
\dtinv f(t)P\bfG&=&
 \int_{t_0}^tdsf(s)P\bfG+\int_{t_0}^tds\int_{t_0}^{s}ds'f(s')
     \{\la\tilde{\bfU}_2, \bfG\ra\bfU_1+\la\tilde{\bfU}_1, \bfG\ra\bfU_2\}, 
     \nonumber \\
                & &+\int_{t_0}^tds\int_{t_0}^{s}ds_1\int_{t_0}^{s_1}ds_2f(s_2)
     \la\tilde{\bfU}_1, \bfG\ra\bfU_1,
\eeq
where the adjoints satisfy 
\beq
A^{\dag}\tilde{\bfU}_1=0,\quad A^{\dag}\tilde{\bfU}_2=\tilde{\bfU}_1,
 \quad A^{\dag}\tilde{\bfU}_3=\tilde{\bfU}_2.
\eeq
The normalization condition reads 
$\la\tilde{\bfU}_3, \bfU_1\ra$
$=\la\tilde{\bfU}_2, \bfU_2\ra=\la\tilde{\bfU}_1, \bfU_3\ra=1$.
\newpage
\setcounter{equation}{0}
\renewcommand{\theequation}{B.\arabic{equation}}
{\large {\bf Appendix B \ \
An elementary method to derive the approximate solution for the
 double-well potential}}\\

The first integral of the Newton equation with 
 the initial condition $x(0)=0$ 
for the double-well potential reads
by
\beq
t=\int_0^{x/\sqrt{2E}}\frac{dy}{1+y^2-\eps Ez^4}.
\eeq
Expanding  the integral in a Taylor series, one readily obtains
\beq
t=(1-\frac{3}{4}\eps E){\rm Sinh} ^{-1}X+
  \eps \frac{ E}{4}\frac{X^3+3X}{\sqrt{X^2+1}},
\eeq
with $X=x/\sqrt{2E}$ up to $O(\eps ^2)$.
Here ${\rm Sinh} ^{-1}X\equiv \ln \vert X+\sqrt{X^2+1}\vert$.
 The point of the present method
is to notice that the above equation is equivalent to the following 
equation up to $O(\eps^2)$;
\beq
\label{dwapp_1}
u={\rm Sinh} ^{-1}X+\frac{\eps E}{4}\frac{X^3+3X}{\sqrt{X^2+1}},
\eeq
with $u=(1+\frac{3}{4}\eps E)t$.
One may solve (\ref{dwapp_1}) perturbatively and obtains
\beq
x(t)=\sqrt{2E}(1-\frac{3}{4}\eps E)\sinh u - 
\frac{\eps }{8}(\sqrt{2E})^3\sinh^3 u.
\eeq
Putting $\sqrt{2E}\{(1-\frac{3}{4}\eps E)=C$, one ends up with
\beq
x(t)=C\sinh \alpha t - \frac{\eps }{8}C^3\sinh^3 \alpha t,
\eeq
with $\alpha =1+\frac{3}{8}\eps C^2$ up to $O(\eps ^2)$,
 which  coincides with 
the result given in subsection 3.3 in the text.

\vspace{1cm}
\newpage
\setcounter{equation}{0}
\renewcommand{\theequation}{C.\arabic{equation}}
{\large {\bf Appendix C \ \
The period of the Lotka-Volterra equation}}
\vspace{1cm}

The Lotka-Volterra equation admit periodic solutions.
An approximate but globally valid periodic solution was explicitly 
constructed
 by the RG method by one of the present author.\cite{kuni97}
The main purpose of this Appendix is to  show that
 the solution constructed  there gives the period which  coincides
  with that obtained 
 by Frame\cite{frame} in a quite different approach.
To make the argument self-contained, however, we shall repeat the RG analysis
 but in a mathematically more simple  way than that
  given in \cite{kuni97}.

Introducing the new variable $\bfu=\ ^t(\xi, \eta)$ by
\beq
x=(b+ \eps\xi)/\eps', \ \ \ \ y=a/\eps + \eta,
\eeq
 Eq.(\ref{eqn:3-1}) is  reduced to the following one:
\beq
\label{eq:lv-2}
\biggl(\frac{d}{dt}- L_0\biggl)\bfu= -\eps\xi\eta\pmatrix{\ 1\cr -1},
\ \ \ \ 
\eeq
where  
\beq
\bfu = \pmatrix{\xi\cr \eta},\ \ \ \ L_0=\pmatrix{0 & -b\cr a & \ 0}.
\eeq
$L_0$ has the eigenvalues
$\lambda =\pm i\sqrt{ab}\equiv \pm i\omega$,
with the corresponding eigenvectors given by
$\bfU _1=\ ^t(1, - i\omega/b)$ and 
$\bfU _2=\bfU^{*}_1$,
respectively. Here $A^{*}$ denotes the complex conjugate of
$A$.
To make the following calculation as transparent as possible,
 we first transform the equation to the form where  $L_0$ is
  diagonalized.
Then one finds that 
the vector equation (\ref{eq:lv-2}) is reduced to a scalar equation
\begin{equation}\label{eqn:3-52}
\left(\frac{d}{dt}-i\omega\right)z
=
\epsilon\frac{i\omega\alpha}{b}(z^2-{\bar z}^2),
\end{equation}
where
$\alpha =1/2\cdot(1-ib/\omega)$ and 
$z(t)=1/2\cdot(\xi+ib/\omega\cdot\eta)$.
This equation apparently has a much simpler form than that treated in 
 \cite{kuni97}.

Now we try to solve (\ref{eqn:3-52}) around $t\sim \forall t_0$ 
by the perturbation theory by expanding
$z=z_0+\epsilon z_1+\epsilon^2z_2+o(\epsilon^2)$,
with the initial condition 
$z(t;t_0)=W(t_0)$.
$W(t_0)$ is also expanded as
$W(t_0)=W_0(t_0)+\eps W_1(t_0)+\eps^2W_2(t_0)+o(\eps^2)$.

A simple manipulation gives
the solution as follows;
\begin{eqnarray}
z(t; t_0)&=&
C(t_0)e^{i\omega t}+
\eps\frac{\alpha}{b}
\left(C^2e^{2i\omega t}+\frac13\bar{C}^{2}e^{-2i\omega t}\right)\nonumber \\ 
 & & +\eps^2\frac{\alpha^2}{b^2}
\left(C^3e^{3i\omega t}-\frac13C\bar{C}^{2}e^{-i\omega t}\right)\nonumber\\
 & &-
\eps^2\frac{|\alpha|^2}{b^2}
\left\{
-\frac12\bar{C}^{3}e^{-3i\omega t}
+\frac{2i\omega}{3}\bar{C}C^2(t-t_0)e^{i\omega t}\right\}.
\label{eqn:3-59}
\end{eqnarray}

Now the RG equation
$\frac{\partial z}{\partial t_0}(t;t_0)\bigm|_{t_0=t}=0$
 gives 
\begin{equation}\label{eqn:3-63}
\frac{dC}{dt_0}
=
-i\frac{\epsilon^2}{6b^2}(1+\frac{b^2}{\omega^2})\omega|C|^2C.
\end{equation}
Here we have used that $|\alpha|^2=\frac14(1+\frac{b^2}{\omega^2})$.
The general solution $z(t, t_0)$ is now given as the initial value $W(t)$
 by construction
$z(t) \equiv W(t)$.

Since  $\vert C(t)\vert^2=$ const$.$ as easily verified from 
(\ref{eqn:3-63}),
one may put $C=\frac{A}{2i}e^{i\theta}$,
 with $A$ being a real constant.
Then (\ref{eqn:3-63}) implies that
$\dot{\theta}=-\epsilon^2A^2/24b^2\cdot(1+b^2/\omega^2)\omega$,
hence
$\theta(t)=-\epsilon^2A^2/24b^2\cdot(1+b^2/\omega^2)\omega t+\bar{\theta}$.

For a calculational convenience, let us define $\Theta$ by
$\Theta(t)=\tilde{\omega}t +\bar{\theta}$ with
\beq
\tilde{\omega}=\left\{
1-\frac{\epsilon^2A^2}{24b^2}(1+\frac{b^2}{\omega^2})
\right\}
\omega ,
\eeq
which implies that 
$C(t)\e^{i\omega t}=A/2\cdot(\sin\Theta-i\cos\Theta).$ 
Then we have finally 
 the components $\ ^t(\xi (t), \eta (t))$, 
\begin{eqnarray}
\xi (t)&=&
A(1-\epsilon^2\frac{\omega^2-b^2}{4\omega^2b^2}\frac{A^2}{12})\sin\Theta 
-
\epsilon^2\frac{1}{b\omega}\frac{A^3}{24}\cos\Theta
-
\epsilon\frac{1}{\omega}\frac{A^2}{6}\sin2\Theta
-
\epsilon\frac{1}{b}\frac{A^2}{3}\cos2\Theta \nonumber \\
&-& 
\epsilon^2\frac{3\omega^2-b^2}{4\omega^2b^2}\frac{A^3}{8}\sin3\Theta
+
\epsilon^2\frac{1}{\omega b}\frac{A^3}{8}\cos3\Theta + o(\epsilon^2), 
\label{eqn:3-70} \\
\nonumber\\
\eta (t)&=& \frac{\omega}{b} 
\{
\epsilon^2\frac{1}{b\omega}\frac{A^3}{24}\sin\Theta
-
A(1+\epsilon^2\frac{\omega^2-b^2}{4\omega^2b^2}\frac{A^2}{12})\cos\Theta
-
\epsilon\frac{1}{b}\frac{A^2}{6}\sin2\Theta
+
\epsilon\frac{1}{\omega}\frac{A^2}{3}\cos2\Theta \nonumber \\
&+&
\epsilon^2\frac{1}{\omega b}\frac{A^3}{8}\sin3\Theta
+
\epsilon^2\frac{\omega^2-3b^2}{4\omega^2b^2}\frac{A^3}{8}\cos3\Theta 
\} + o(\epsilon^2). \label{eqn:3-71}
\end{eqnarray}
Although these expressions are seemingly different from those given in \cite
{kuni97}, they coincides with each other up to $o(\eps^2)$.
In fact, by the redefinition of the constant variables
$A(1-\epsilon^2\frac{\omega^2-b^2}{4\omega^2b^2}\frac{A^2}{12})$ 
  $\rightarrow  A$ and
$\bar{\theta}-\eps^2\frac{1}{b\omega}\frac{A^3}{24}$
 $\rightarrow \bar{\theta}$,
(\ref{eqn:3-70}) and (\ref{eqn:3-71}) reproduce  the result
 in \cite{kuni97}, up to $o(\eps ^2)$.

It is remarkable that we have obtained the modified angular velocity 
$\tilde{\omega}$ depending on the amplitude.
A long ago, Frame\cite{frame} gave the period  of the motion. 
Here, we shall show that the period $T_{\rm RG}=2\pi/\tilde{\omega}$ 
coincides with that given by Frame up to o$(\eps^2)$.

Our  period reads
\begin{equation}\label{eqn:3-74}
T_{\rm RG}    = \frac{2\pi}
{ \omega-\frac{1}{\omega}\frac{\epsilon^2A^2}{24b^2}(\omega^2+b^2) }
    = \frac{2\pi}{\sqrt{ab}}
\{1+\frac{\epsilon^2A^2}{24b}(\frac 1a+\frac 1b)\}+o(\epsilon^{2}),
\end{equation}
where use has been made of $\omega=\sqrt{ab}$.
On the other hand, the period $T_{\rm Fr}$ given by Frame \cite{frame}
 reads in his notations
\begin{equation}\label{eqn:3-76}
 T_{\rm Fr} = \frac{2\pi}{a_1a_2}
\{
1+\frac{(c_1c)^2}{1!1!}+\frac{(c_1c)^4}{2!2!}+\cdots
\}.
\end{equation}
Here,  $c_1, c$ are identified with the variables expressed by
 our basic variables as follows
;
\begin{eqnarray}
c_1^2&=&\frac{a+b}{24},\label{eqn:3-78} \\
c^2&=&\frac 2a
\{ \frac{\epsilon}b\xi-\log(1+\frac{\epsilon}b\xi) \}
+
\frac 2b
\{ \frac{\epsilon}a\eta-\log(1+\frac{\epsilon}a\eta) \} \label{eqn:3-79}.
\end{eqnarray}
We remark that
the correspondence between the variables of ours and Frame's is
 summarized as follows:
\begin{eqnarray}
& &a\leftrightarrow a_1^2 \ ;\ 
b\leftrightarrow a_2^2 \ ;\ 
\epsilon\leftrightarrow a_{12} \ ;\ 
\epsilon'\leftrightarrow a_{21} \nonumber\\
& &x\leftrightarrow N_1 \ ;\ 
y\leftrightarrow N_2 \ ;\ 
\frac b{\epsilon'} \leftrightarrow n_1 \ ;\ 
\frac a{\epsilon} \leftrightarrow n_2. \nonumber
\end{eqnarray}
It should be remarked here that $c^2$ is actually a constant because
 our system has a conserved quantity
$b \ln x-\eps'x +a \ln y -\eps y={\rm const}$.
In fact, one can easily verify that
\begin{equation}\label{eqn:3-80}
c^2
=
\frac{\epsilon^2}{ab}
(\frac{\xi^2}b+\frac{\eta^2}a)
+0(\epsilon^2).
\end{equation}
Thus,
\beq
\label{eq:period}
T_{\rm Fr}
=
\frac{2\pi}{\sqrt{ab}}
\{
1+c_1^2c^2
\}
+0(\epsilon^2)
=
\frac{2\pi}{\sqrt{ab}}
\{
1+\frac{a+b}{24}
\frac{\epsilon^2}{ab}
(\frac{\xi^2}b+\frac{\eta^2}a)
\}
+o(\epsilon^2).
\eeq
 From (\ref{eqn:3-70})
 and (\ref{eqn:3-71}), one sees that
$\xi  =0+o(1),\quad \eta =\frac{\omega}{b}A+o(1)$.
Hence,

\begin{eqnarray}
T_{\rm Fr} &=& 
\frac{2\pi}{\sqrt{ab}}
\{
1+\frac{\epsilon^2}{24}(\frac 1a+\frac 1b)\frac{A^2}b
\}
+o(\epsilon^2) \nonumber \\
 \ &=& T_{\rm RG}+o(\eps^2).\label{eqn:3-81}
\end{eqnarray}
 This is what we wanted to show.

\end{document}